\documentclass[eqsecnum,pra,aps]{revtex4}

\usepackage{graphicx,amssymb,bm}
\usepackage{color}

\def\ket#1{|\,#1\,\rangle}
\def\bra#1{\langle\, #1\,|}

\def\proj#1#2{\ket{#1}\bra{#2}}

\newcommand{\Ignore}[1]{ }

\newcommand{\beq}{\begin{equation}}
\newcommand{\eeq}{\end{equation}}
\newcommand{\beqa}{\begin{eqnarray}}
\newcommand{\eeqa}{\end{eqnarray}}
\newcommand{\average}[1]{\langle {#1} \rangle}

\begin{document}

\title{Quantum Correlation Dynamics in Controlled Two-Coupled-Qubit Systems}

\author{Iulia Ghiu$^1$}

\author{Roberto Grimaudo$^{2}$}

\author{Tatiana Mihaescu$^{3,4}$}

\author{Aurelian Isar$^4$}

\author{Antonino Messina$^{5}$}

\affiliation{$^1$University of Bucharest, Faculty of Physics, Centre for Advanced Quantum Physics, PO Box MG-11, R-077125, Bucharest-Magurele, Romania}

\affiliation{$^2$Dipartimento di Fisica e Chimica dell'Universit\`a di Palermo, Via Archirafi 36, I-90123 Palermo, Italy}

\affiliation{$^3$Faculty of Physics, University of Bucharest, POB MG-11, Bucharest-Magurele, Romania}

\affiliation{$^4$Department of Theoretical Physics, National Institute of Physics and Nuclear Engineering, POB MG-6, Bucharest-Magurele, Romania}

\affiliation{$^5$Dipartimento di Matematica ed Informatica dell'Universit\`a di Palermo, Via Archirafi 34, I-90123 Palermo, Italy }

\begin{abstract}
We study and compare the time evolutions of concurrence and quantum discord in a driven system of two interacting qubits prepared in a generic Werner state. The~corresponding quantum dynamics is exactly treated and manifests the appearance and disappearance of entanglement. Our analytical treatment transparently unveils the physical reasons for the occurrence of such a phenomenon, relating it to the dynamical invariance of the $X$ structure of the initial state. The~quantum correlations which asymptotically emerge in the system are investigated in detail in terms of the time evolution of the fidelity of the initial Werner state.
\end{abstract}

\maketitle

\section{Introduction}

Entanglement sudden death is a phenomenon that was widely investigated during recent years in the case of open quantum systems~\cite{Eberly-2009} and represents the decrease of the entanglement to zero in a finite time. For example, the entanglement sudden death was studied in the following quantum systems: two atoms locally coupled to the modes of their cavities~\cite{Eberly-jpb-2007}, two qubits in an $X$ state under the action of phase damping, amplitude damping, bistability noise~\cite{Eberly-2007}, polarization-entangled photon pairs under the influence of polarization mode dispersion~\cite{Brodsky}, or amplitude damping channel~\cite{Singh}. On~the other hand, the opposite concept, i.e., sudden birth of entanglement was considered in Ref.~\cite{Ficek} by using the dissipative process of spontaneous emission and in Ref.~\cite{Bellomo}, where the dynamics of two quantum emitters, which interact with a stationary electromagnetic field out of thermal equilibrium, is~in detail explained.

Over recent years a lot of attention has been devoted to the analysis of the combined process, i.e., sudden death, followed by revival of entanglement. These two linked phenomena have been investigated in the case of two qubits interacting with a common reservoir~\cite{Mazzola,Namitha}, trapped atoms or ions under the influence of applied pulses~\cite{Sola}, two cavities interacting with independent reservoirs~\cite{Lopez}, quantum systems subjected to a classical random external field~\cite{Metwally}, two-level atoms in the presence of a single mode quantized field~\cite{Bahari}, dark-soliton qubits~\cite{Shaukat}, a diamond sample interacting with a solid-state spin bath~\cite{Wang}.

In this paper, we investigate the quantum dynamics of two interacting qubits, each one subjected to a local time-dependent magnetic field. Our main goal is to verify the occurrence of sudden death and rebirth manifestations in the time evolution of the quantum correlations arising between the two qubits in such a controlled, time-dependent, physical scenario. To this end we investigate the time dependence of the  concurrence to reveal the presence of entanglement between the two qubits. Since, however, to know that the system is in a separable state does not preclude the possible existence of nonclassical correlations in this state, in this paper we go beyond the concurrence. Indeed we exactly evaluate in which way the quantum discord goes with time, since it captures all kinds of nonclassical correlations (entanglement included) and  then can be different from zero even when there is no entanglement.

Quantum discord (QD) is defined in Ref.~\cite{Zurek} as the difference between the quantum generalizations of two equivalent classical expressions of the mutual information.
It is of interest to highlight that the quantum discord possessed by a bipartite system, especially when it is in a separable state, is today considered a possible resource for the development of quantum technologies, especially in the quantum computation field~\cite{Lanyon,Datta,Guo}. In other words, the idea that the presence of quantum correlations necessarily requires the existence of entanglement must be considered wrong~\cite{Meyer}. Thus, given that finding nonclassical correlations in a composite quantum system would provide a strategic key to improving the yield of the quantum information processing, the study of quantum discord has received a great impulse in recent years~\cite{revmodi,kmodi,disc2,disc3}.
In general, unfortunately, the analytical formula of quantum discord is difficult to be obtained, since it requires an extremization procedure.
The~reason making computing QD so difficult stems from the fact that the time required for such a target becomes exponentially larger and larger as a function of the dimension of the Hilbert space of the bipartite system under scrutiny~\cite{Huang}.
In the case of continuous variable systems, for example Gaussian states, an explicit formula of quantum discord was however found, if one restricts the set of all quantum measurements to Gaussian ones~\cite{Paris-2010,Adesso-2010}. A comparison between discord and entanglement of a two-mode Gaussian state, as well as the study of non-Gaussianity under the influence of local baths was made in Refs.~\cite{isar1,Marian,Isar,Ghiu-2014}. On~the other hand, for discrete quantum systems such as for two qubits, the characterization of quantum discord is difficult to be made in the general case. For the particular situation of the so-called class of two-qubit $X$ states, the quantum discord was evaluated first numerically~\cite{Maziero}, and then analytically~\cite{Ali,Li}. To exploit such a result in the study of the quantum dynamics of our system, it is necessary to prepare it in a mixed $X$ state. To this end in this paper we assume that the initial state of the system is a generic Werner state~\cite{werner}. It is of relevance that quite recently a new method for synthetizing and characterizing these states have been reported~\cite{barb,cine}. Werner states are $X$ states exhibiting intriguing nonlocal correlations~\cite{popes} and, therefore, play an important role in the quantum information processing. All these features make of special interest investigating their dynamics when the physical scenario evolves under controlled time-dependent~fields.

Our aim in this paper is two-fold: on one hand, we show that sudden death, followed by revival of entanglement, occurs in the case of two-spin-1/2 particles in the presence of time-dependent magnetic fields. On~the other hand, a detailed comparison between the dynamics of concurrence and quantum discord is made, carefully dwelling on those situations when the mixed state is separable, but is described by non-zero quantum discord.
Over recent years many papers comparing concurrence and quantum discord in a 4-dimensional Hilbert space have been published. Some of them~\cite{Ali,galve,rau,virzi,Castro-2018} report this comparison in an appropriate space of the real parameters characterizing a priori selected families of mixed states. Others, instead, show the dynamical evolution of concurrence and quantum discord, generated in a chosen open and time-independent physical scenario~\cite{fanchini,auyu,liu1,gallego,aiobi,park,egel}. At the best of our knowledge only one paper analyzing the time evolution of concurrence versus that of quantum discord, generated by a time-dependent Hamiltonian---when the system is prepared in a convex combination of two Bell states, has been published so far~\cite{Roberto}. The~analytical solutions found
for the Hamiltonian model given in Ref.~\cite{Messina-2016} constitutes the platform on which the exact evaluation of
both concurrence and quantum discord on this paper is based.

The paper is organized as follows. In Section~\ref{canonic} we review the concept of the canonical form of $X$ states. This section is quite important, since all the mixed states used in this paper are $X$ states and the evaluation of quantum discord is based on transforming an arbitrary $X$ state to its canonical form. Section~\ref{hamilt-model} is devoted to the Hamiltonian model of two-spin-1/2 particles subjected to time-dependent magnetic fields. We show that the initial $X$ structure of the density operator is preserved during the evolution of the system under scrutiny. A detailed analysis of the dynamics of the concurrence and quantum discord, including a comparison between their behaviors, is presented in Section~\ref{dynamics} by considering that at the initial time the density operator of the two qubits is a Werner state. A special class of one real parameter two-qubit states, which represents an extension of that of Werner states, is constructed easily showing that the separability condition of the Werner initial state still holds for the evolved states too. By using the approach of Li~\cite{Li} presented in Appendix~\ref{sec-discord}, we compute the quantum discord of the two qubits subjected to magnetic fields. In addition, we show that sudden death, followed by revival of entanglement, occurs for some interval of values for the parameter $\alpha$ which characterizes the initial Werner state. Furthermore, we present in Section~\ref{sec-fid} an explanation of the asymptotic behavior of the two measures of correlations studied in the previous sections, i.e., concurrence and quantum discord. This interpretation is based on the time evolution of the fidelity between the initial Werner state and the evolved Werner state.
Our conclusions are drawn in Section~\ref{sec-concl}. The~exact solutions of the two-spin-1/2 particles described by the Hamiltonian model of Section~\ref{hamilt-model} are presented in detail in Appendix~\ref{exact-solutions}. Appendix~\ref{constant} is devoted to the analysis of the behavior of both concurrence and quantum discord when constant magnetic fields are applied. The~analytical expression of the fidelity between the Werner state and the generalized Werner state is obtained in Appendix~\ref{Fid W States}.

\section{Preliminaries: Canonical Form of $X$ States}
\label{canonic}

The Bloch generalization of the density operator of a qubit to the case of two-qubit systems is given by the parametrization introduced by Fano~\cite{Fano}. The~general expression of a two-qubit density operator acting in the Hilbert space ${\cal H}_A\otimes {\cal H}_B$ is~\cite{Fano,Horodecki}:
\begin{equation}
\rho = \frac{1}{4}\, \left( I\otimes I + {\bf r}\cdot {\boldsymbol \sigma} \otimes I+ I \otimes {\bf s}\cdot {\boldsymbol \sigma} +\sum_{m,n=1}^3t_{mn}\, \sigma_m \otimes \sigma_n \right),
\label{op-gen}
\end{equation}
where $\sigma_j$, with $j$ = 1, 2, 3 are the Pauli operators. Equation (\ref{op-gen}) represents the Fano parametrization of $\rho$.
The vectors ${\bf r}$ and ${\bf s}$ are real, their expressions being $r_j=\mbox{Tr}(\rho \, \sigma_j\otimes I)$ and $s_j=\mbox{Tr}(\rho \, I\otimes \sigma_j)$. The~matrix $T$ defined by $t_{mn}$ is a real matrix, with $t_{mn}=\mbox{Tr}(\rho \, \sigma_m\otimes \sigma_n)$, where $m,n$ = 1, 2, 3.

Let us briefly discuss the transformation of a two-qubit density operator under a local unitary transformation. For any single-qubit unitary transformation $U$ there is a unique rotation $O$ such that:
\begin{equation}
U\, {\bf n}\cdot {\boldsymbol \sigma} \, U^\dagger = (O\, {\bf n})\cdot {\boldsymbol \sigma} .
\label{unit-rot}
\end{equation}

Let us denote by $\tilde \rho$ the transformed density operator obtained by applying a local unitary transformation $U_A\otimes U_B$:
$
\tilde \rho =U_A\otimes U_B \, \rho \, U_A^\dagger \otimes U_B^\dagger .
$
Hence, the parameters ${\bf r}$, ${\bf s}$, and $T$ transform as~\cite{Horodecki}:
\begin{eqnarray}
{\bf {\tilde r}}&=&O_A\, {\bf r}; \; \;  \; \; {\bf {\tilde s}}=O_B\, {\bf s}, \nonumber \\
\tilde T&=&O_A\, T\, O_B^T \label{transf-T},
\end{eqnarray}
where $O_A$ and $O_B$ are related to $U_A$ and $U_B$, respectively, through Equation (\ref{unit-rot}).

A widely studied family of two-qubit states is the so-called class of $X$ states, whose density operator is characterized by non-vanishing entries only along the diagonal and the anti-diagonal:
\begin{equation}
\rho_{\mbox x}=\left(
\begin{array}{cccc}
 \rho_{11} & 0 & 0 & \rho_{14} \\
 0 & \rho_{22} & \rho_{23} & 0 \\
 0 & \rho_{32} & \rho_{33} & 0 \\
 \rho_{41} & 0 & 0 & \rho_{44} \\
\end{array}
\right),
\label{x-st}
\end{equation}
where $\rho_{jj}$ are real, with $j$ = 1, 2, 3, 4, while the off-diagonal terms are complex.
Let us denote $\rho_{14}=|\rho_{14}|\, e^{i\, \varphi_{14}}$ and $\rho_{23}=|\rho_{23}|\, e^{i\, \varphi_{23}}$. In addition, one has $\rho_{41}=\rho_{14}^*$ and $\rho_{32}=\rho_{23}^*$. The~unit trace condition is given by $\sum_{j=1}^4 \rho_{jj}=1$, while the positivity condition reads $\rho_{11}\rho_{44}\geq|\rho_{14}|^2$ and $\rho_{22}\rho_{33}\geq|\rho_{23}|^2$. All the matrices are represented in this paper in the ordered computational basis $\{\ket{00},\ket{01},\ket{10},\ket{11}\}$.
The Fano parametrization of an $X$ state is given by:
\begin{eqnarray}
{\bf r_{\mbox x}}:&& 0, 0, r; \nonumber \\
{\bf s_{\mbox x}}:&& 0, 0, s; \label{fano-x} \\
T_{\mbox x}&=&\left( \begin{array}{ccc}
T_{11}&T_{12}&0\\
T_{21}&T_{22}&0\\
0&0&T_{33}
\end{array} \right). \nonumber
\end{eqnarray}

The link between the general form (\ref{x-st}) and its Fano parametrization (\ref{fano-x}) is given by~\cite{Rau-2009}:
\begin{eqnarray}
r&=& \rho_{11}+\rho_{22} -\rho_{33}-\rho_{44}, \nonumber\\
s&=& \rho_{11}-\rho_{22}+ \rho_{33}-\rho_{44},  \nonumber\\
T_{11}&=&2\, {\rm Re}[\rho_{23}+\rho_{14}], \nonumber\\
T_{22}&=&2\, {\rm Re}[\rho_{23}-\rho_{14}], \nonumber\\
T_{33}&=&\rho_{11}-\rho_{22}- \rho_{33}+\rho_{44}, \nonumber \\
T_{12}&=&2\, {\rm Im}[\rho_{23}-\rho_{14}],\nonumber\\
T_{21}&=&-2\, {\rm Im}[\rho_{23}+\rho_{14}].\nonumber
\end{eqnarray}

One can diagonalize $T$ by applying two rotations $O_A$ and $O_B$ along the $Ox_3$-axis, associated with the following local unitary operation, according to Equations (\ref{unit-rot}), (\ref{transf-T})~\cite{Simon-2013,Huang-2013,Yuri-2015,Celeri-2017}:
\begin{equation}
\tilde U_A\otimes \tilde U_B=e^{-i\, (\varphi_{14}+\varphi_{23})\, \sigma_3/4}\otimes e^{-i\, (\varphi_{14}-\varphi_{23})\, \sigma_3/4}.
\label{op-unit-loc}
\end{equation}

{\it The canonical form} of a general $X$ state is $\rho_{\mbox x}^{can} =\tilde U_A\otimes \tilde U_B \, \rho_{\mbox x} \, \tilde U_A^\dagger \otimes \tilde U_B^\dagger $~\cite{Simon-2013}:
\begin{equation}
\rho_{\mbox x}^{can}=\left(
\begin{array}{cccc}
 \rho_{11} & 0 & 0 & |\rho_{14}| \\
 0 & \rho_{22} & |\rho_{23}| & 0 \\
 0 & |\rho_{32}| & \rho_{33} & 0 \\
 |\rho_{41}| & 0 & 0 & \rho_{44} \\
\end{array}
\right).
\label{x-can}
\end{equation}

The Fano parametrization of the canonical form of the $X$ state (\ref{x-can}) is given by $T$ = diag($c_1,c_2,c_3$):
\begin{eqnarray}
r^{can}&=& r = \rho_{11}+\rho_{22} -\rho_{33}-\rho_{44},\nonumber \\
s^{can}&=& s = \rho_{11}-\rho_{22}+ \rho_{33}-\rho_{44}, \nonumber \\
c_1&=&T_{11}^{can}=2\, (|\rho_{23}|+|\rho_{14}|),\label{fano-can}\\
c_2&=&T_{22}^{can}=2\, (|\rho_{23}|-|\rho_{14}|),\nonumber\\
c_3&=&T_{33}^{can}=T_{33} = \rho_{11}-\rho_{22}- \rho_{33}+\rho_{44}. \nonumber
\end{eqnarray}

Therefore, the canonical form of the Fano parametrization of the density operator of an $X$ state is given by:
\begin{equation}
\rho_{\mbox x}^{can} = \frac{1}{4}\, \left( I\otimes I + r\, \sigma_3 \otimes I+ s\, I \otimes \, \sigma_3 +\sum_{j=1}^3c_j\, \sigma_j \otimes \sigma_j \right).
\label{x-can-fano}
\end{equation}

Since the quantum correlations remain invariant under local unitary transformations, the method of bringing an arbitrary $X$ state to its canonical form is of great importance. A deep understanding of the description of the canonical form of the Fano parametrization of an $X$ state is crucial for evaluating different measures of quantum correlations.
To compute the quantum discord of some specific $X$ states, we will use the approach presented here in Section~\ref{dynamics}.

\section{Time-Dependent Hamiltonian Model and the Related Evolution Operator}
\label{hamilt-model}

Consider a two-spin-1/2 system under the influence of two time-dependent magnetic fields ${\bf B}_k(t)=(0,0,B_k(t))$, where $k=A,B$. We denote by $g_A$ and $g_B$ the real, positive, dimensionless coefficients that contain the corrections to the coupling terms between each spin and the local magnetic field applied on it. One can define~\cite{Messina-2016}:
$$\omega_k(t)=\frac{1}{2}\, \mu_B\, g_k\, B_k(t), $$
where $k=A, B$. The~two-spin-1/2 Hamiltonian model we discuss here has been investigated in Ref.~\cite{Messina-2016}:
\begin{eqnarray} \label{Hamiltonian}
H=
\hbar\omega_A\sigma_3\otimes I+\hbar\omega_BI\otimes \sigma_3+\gamma_{11}\sigma_1\otimes \sigma_1+\gamma_{22}
\sigma_2\otimes \sigma_2+\gamma_{33}\sigma_3\otimes \sigma_3+\gamma_{12}\sigma_1\otimes
\sigma_2+\gamma_{21}\sigma_2\otimes \sigma_1.
\end{eqnarray}

Such a model has been used~\cite{Messina-2016} to describe two interacting spin-1/2's subjected to local, generally time-dependent, magnetic fields [$\omega_1(t)$ and $\omega_2(t)$], while the coupling parameters are intended to be time-independent.
The first three interaction terms account for anisotropic Heisenberg interaction, while the last two terms stem from asymmetric dipole-dipole~\cite{Bolton} and Dzyaloshinskii-Moriya~\cite{Dzyaloshinskii,Moriya} interactions.

In Ref.~\cite{Messina-2016}, it has been proved that as a consequence of the symmetry properties of $H$, the time evolution operator, solution of the Schr\"odinger equation $i\hbar\dot{U}=HU$, keeps the following $X$ structure at any time
\begin{equation}
U(t)=\left( \begin{array}{cccc}
a_+ & 0 & 0 & b_+ \\
0 & a_- & b_- & 0 \\
0 & -b_-^* & a_-^* & 0 \\
-b_+^* & 0 & 0 & a_+^* \\
\end{array} \right), ~~~a_\pm(t)\equiv|a_\pm(t)|e^{i\phi_a^\pm(t)}, ~~~ b_\pm(t)\equiv|b_\pm(t)|e^{i\phi_b^\pm(t)},\label{unit-U}
\end{equation}
where the parameters $a_\pm(t)$ and $b_\pm(t)$, in general, depend on the Hamiltonian parameters.

Since $U(0)=I\otimes I$, then $a_{\pm}(0)=1$ and $b_{\pm}(0)=0$. In addition,
it has been shown~\cite{Messina-2016} that the $2 \times 2$ unitary operators
\begin{displaymath}
U_{\pm}=e^{\mp i \gamma_{33} t / \hbar}
\left( \begin{array}{cc}
a_\pm & b_\pm \\
-b_\pm^* & a_\pm^*
\end{array} \right)
\end{displaymath}
are the time evolution operators generated by the following single spin-1/2 Hamiltonians
\begin{displaymath}
H_{\pm}=\left(
\begin{array}{cc}
\Omega_{\pm} & \Gamma_{\pm} \\
\Gamma_{\pm}^{*} & -\Omega_{\pm}
\end{array}  \right)
\pm \gamma_{33} I,
\end{displaymath}
where
\begin{equation}
\Omega_{\pm}(t) = \hbar [ \omega_A \pm \omega_B ], \qquad
\Gamma_{\pm} = (\gamma_{11} \mp \gamma_{22}) - i (\pm \gamma_{12} + \gamma_{21}).
\label{omega-gama}
\end{equation}

An interesting dynamical property of the Hamiltonian model consists of the fact that the $X$ structure of an initial state is preserved during the evolution~\cite{Roberto}.
Indeed, suppose that the two-spin-1/2 system is initially prepared in a general $X$ state, as given by Equation (\ref{x-st}).
The non-zero entries of the $X$-state $\rho(t)=U(t)\rho_{\mbox x}(0)U^\dagger(t)$ may be expressed as follows
\begin{eqnarray}
\rho_{11}(t)&=& |a_+|^2\rho_{11}+|b_+|^2\rho_{44}+2\, \text{Re}[a_+b_+^*\rho_{14}] \nonumber \\
\rho_{14}(t)&=&\rho_{41}^*(t)= a_+^2\rho_{14}-b_+^2\rho_{41}-a_+b_+(\rho_{11}-\rho_{44}) \nonumber \\
\rho_{22}(t)&=& |a_-|^2\rho_{22}+|b_-|^2\rho_{33}+2\, \text{Re}[a_-b_-^*\rho_{23}] \label{ro-t} \\
\rho_{23}(t)&=&\rho_{32}^*(t)= a_-^2\rho_{23}-b_-^2\rho_{32}-a_-b_-(\rho_{22}-\rho_{33}) \nonumber \\
\rho_{33}(t)&=& |b_-|^2\rho_{22}+|a_-|^2\rho_{33}-2\, \text{Re}[a_-b_-^*\rho_{23}] \nonumber \\
\rho_{44}(t)&=& |b_+|^2\rho_{11}+|a_+|^2\rho_{44}-2\, \text{Re}[a_+b_+^*\rho_{14}].\nonumber
\end{eqnarray}

We emphasize that such a dynamical decomposition was successfully used: (1) to bring to light peculiar physical effects like the coupling-based Landau-Zener transitions in the two-qubit system~\cite{GVM1}, as well as (2) to treat and solve the exact dynamics of more complex system like two interacting qutrits~\cite{GMIV,GVM2}, two coupled qubits~\cite{GBNM} and $N$ spin 1/2's coupled through high order interaction terms~\cite{GLSM}.

\section{Dynamics of Concurrence and Quantum Discord of the Evolved Werner State for Time-Dependent Magnetic Fields }
\label{dynamics}

The results mentioned in the previous section may be summarized claiming that the solution of the dynamical problem of the two coupled spin-1/2's may be traced back to the solution of two independent single spin-1/2 dynamical problems~\cite{Messina-2016}.
However, depending on the time-profiles of the two magnetic fields, we might not be able to analytically solve the sub-dynamical problems too.
In~Ref.~\cite{Messina-2016} the following exactly solvable time-dependent scenarios have been proposed:

Case 1. The~two magnetic fields vary over time as follows
\begin{equation}
\hbar \omega_{A,B}(t) = \frac{|\Gamma_+|}{\cosh(2\tau_+)} \pm \frac{|\Gamma_-|}{\cosh(2\tau_-)}.
\label{cimp-caz1}
\end{equation}

Case 2. The~two magnetic fields vary over time as follows
\begin{equation}
\hbar \omega_{A,B}(t) = \frac{|\Gamma_+|}{\cosh(2\tau_+)} \pm
{|\Gamma_-| \over 4} \biggl[ { 3 \over \cosh(\tau_-) } - \cosh(\tau_-) \biggr],
\label{cimp-caz2}
\end{equation}
where we have defined
\begin{equation}
 \tau_\pm:={|\Gamma_\pm| \over \hbar}\, t.
 \label{tau-pm}
\end{equation}

We underline that such cases are just two exactly solvable examples that can be derived by the knowledge of analytical solutions of the single spin-1/2 dynamical problem. Other analytically solvable cases may be constructed based on the solutions reported in Refs.~\cite{Bagrov, Kuna, Barnes, MN, MGMN, GdCNM, SNGM}.

Let us suppose that at the initial time $t=0$ the state of the two-spin-1/2 system is a Werner state~\cite{werner}:
\begin{equation}
\rho_W^{(\alpha )}=\frac{1-\alpha }{4}\, I\otimes I +\alpha \, \proj{\Psi^-}{\Psi^-},
\label{st-werner}
\end{equation}
where $\ket{\Psi^-}=\frac{1}{\sqrt 2}\, (\ket{01}-\ket{10})$ is the singlet state and $\alpha \in [-\frac{1}{3},1]$.
The Werner state (\ref{st-werner}) is a particular $X$ state (\ref{x-st}), being characterized by:
\begin{eqnarray}
\rho_{11}=\rho_{44}&=&\frac{1-\alpha }{4},\nonumber \\
\rho_{22}=\rho_{33}&=&\frac{1+\alpha }{4},\nonumber \\
\rho_{23}=\rho_{32}&=&-\frac{\alpha }{2}, \nonumber
\end{eqnarray}
with all the other entries equal to zero.

By using Equation (\ref{ro-t}) we find the expressions of the non-zero elements of the evolved density matrix $\rho(t)=U(t)\rho_W^{(\alpha )}U^\dagger(t)$, where $U(t)$ is given by Equation (\ref{unit-U}):
\begin{eqnarray}
\rho_{11}(t)&=&\rho_{44}(t)=\frac{1-\alpha}{4},\nonumber \\
\rho_{22}(t)&=&\frac{1+\alpha}{4}\,-\alpha\, \text{Re}[a_-b_-^*], \nonumber \\
\rho_{23}(t)&=&\frac{\alpha}{2}\, (b_-^2-a_-^2)=\rho_{32}^*(t),\label{st-x-timp} \\
\rho_{33}(t)&=&\frac{1+\alpha}{4}\,+\alpha\, \text{Re}[a_-b_-^*]. \nonumber
\end{eqnarray}

An equivalent expression of the evolved state can be written as follows:
\begin{equation}
\rho(t)= \frac{1-\alpha }{4}\, I\otimes I +\alpha \, \proj{\psi(t)}{\psi(t)},
\label{ro-t-st-fi}
\end{equation}
where we have denoted
\begin{equation}
\ket{\psi(t)}=U(t)\, \ket{\Psi^-}=c_{01}(t)\ket{01}+c_{10}(t)\ket{10}.
\label{psi-timp}
\end{equation}

The states (\ref{psi-timp}) belong to the class of the so-called Werner–Popescu states~\cite{popes}, since the evolution operator $U$ cannot be represented as the tensorial product of unitary operators acting in the bidimensional Hilbert spaces of the two qubits.
The time-dependent coefficients $c_{01}(t)$ and $c_{10}(t)$ are given by:
\begin{eqnarray}
c_{01}(t)~=~\frac{1}{\sqrt 2}\, \exp \left( i\, \frac{\gamma_{33}}{\hbar} \, t\right) \, \left( a_- - b_-  \right), \label{c01}\\
c_{10}(t)~=~-\frac{1}{\sqrt 2}\, \exp \left( i\, \frac{\gamma_{33}}{\hbar} \, t\right) \, \left( a_-^* + b_-^* \right). \label{c10}
\end{eqnarray}

In the following sections, we investigate the behavior of the concurrence and quantum discord for the two cases (\ref{cimp-caz1}) and (\ref{cimp-caz2}), with the two qubits initially prepared in the Werner state (\ref{st-werner}).

\subsection{Concurrence}
\label{sec-conc}

To investigate the dynamics of the correlations, we use as a measure of entanglement of the two qubits the concurrence, which was introduced by Wootters~\cite{Wootters-1998, Wootters-2001}. Consider a pure state
$\ket{\phi}= a\, \ket{00} + b\, \ket{01} + c\, \ket{10} + d\, \ket{11}. $
Then the concurrence for such a state is:
\begin{equation}
C(\ket{\phi})=2\, |a\, d-b\, c|.
\label{conc-st-pura}
\end{equation}

If the initial state is the singlet one $\ket{\Psi^-}$, i.e., the Werner state (\ref{st-werner}) with $\alpha =1$, then the evolved state under the influence of the magnetic fields is given by $\ket{\psi (t)}$ (\ref{psi-timp}).

The concurrence has the expression $C(\ket{\psi (t)})=2\, |c_{01}\, c_{10}|$ according to Equation (\ref{conc-st-pura}), where $c_{01}$ and $c_{10}$ are given by Equations (\ref{c01}) and  (\ref{c10}). With the help of Equations (\ref{a6}) and (\ref{a11}) from the Appendix~\ref{exact-solutions}, we find the analytical expressions of the concurrence for the two cases of the applied magnetic fields:

  $C(\ket{\psi (t)})= \sqrt{1-\tanh ^2(2\tau_-)\sin^2(2\tau_-)} $   for the Case 1 of Equation (\ref{cimp-caz1});

\begin{figure}
\centering
\includegraphics[width=7.5cm]{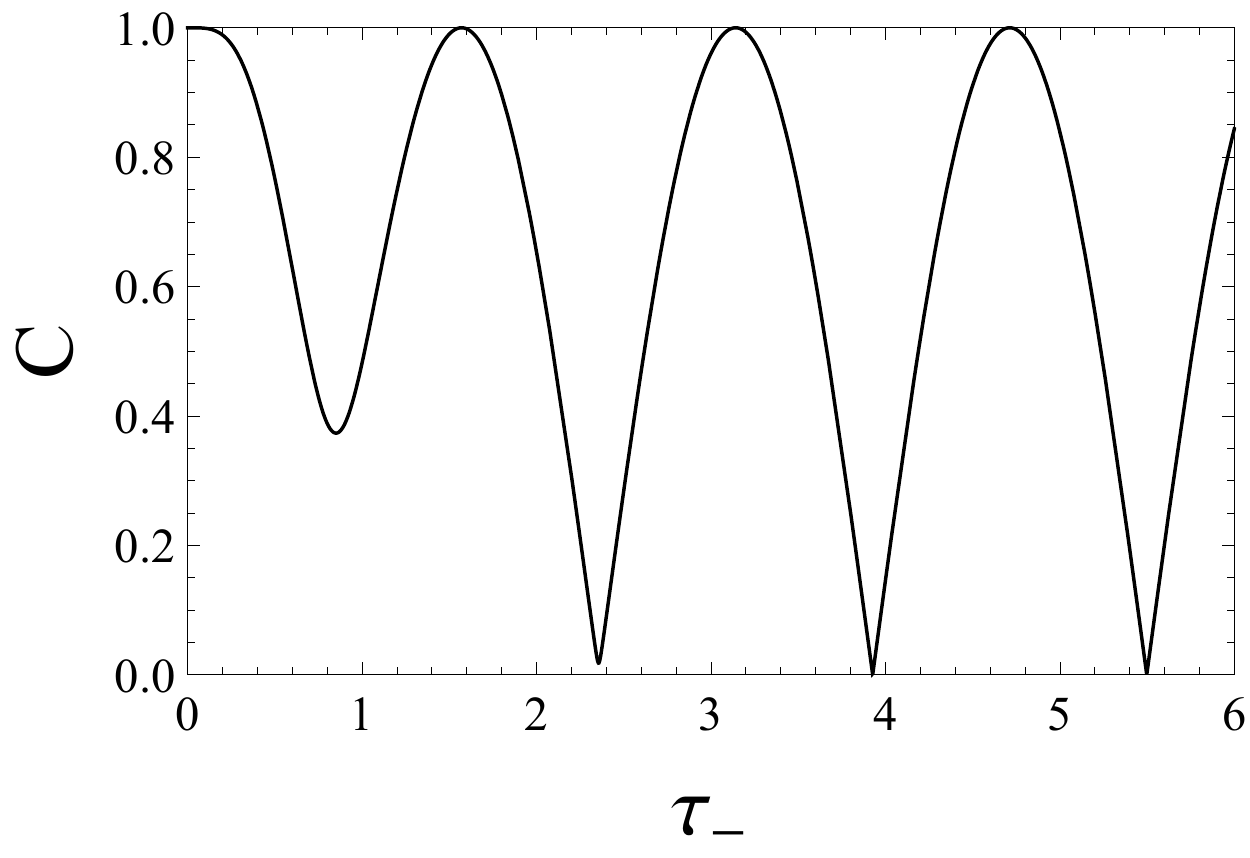}
\includegraphics[width=7.5cm]{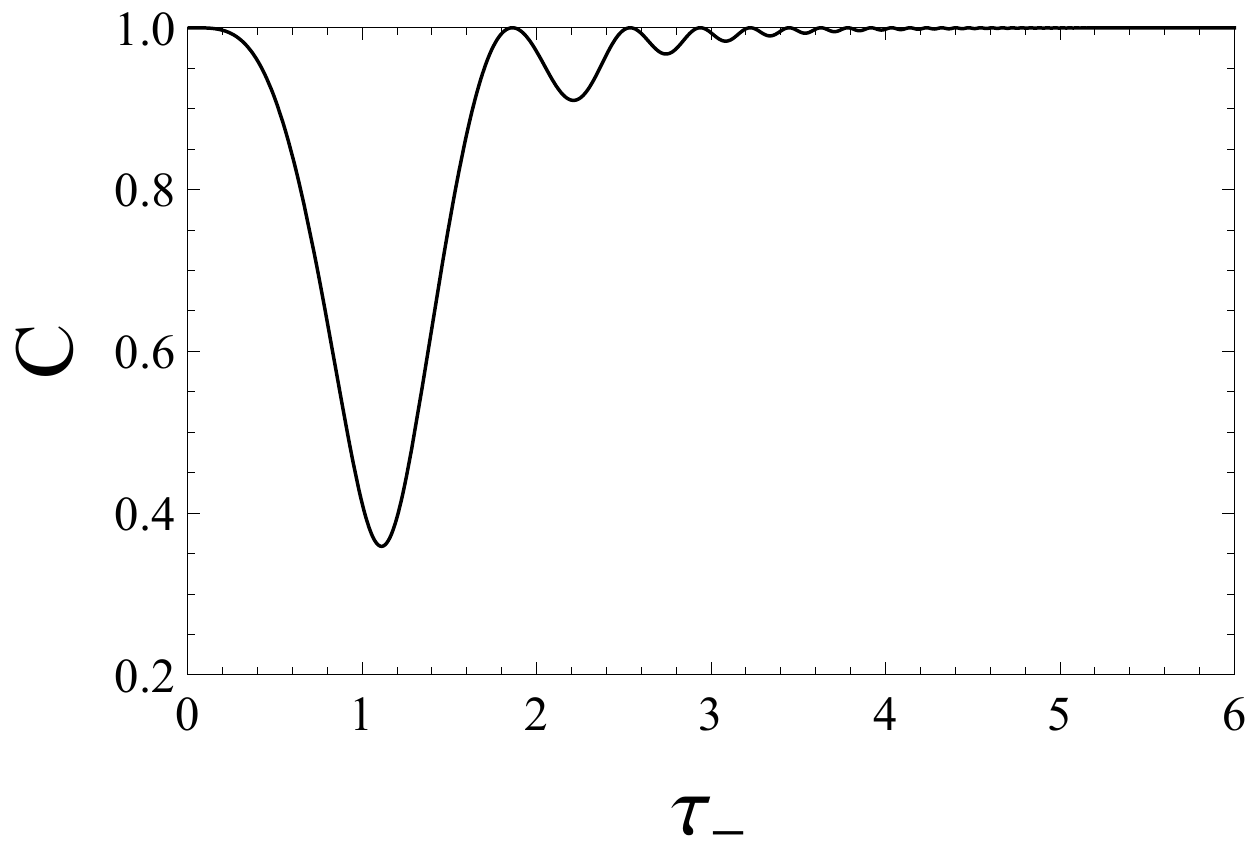}
\caption{Concurrence when the state at $t=0$ is the singlet state $\ket{\Psi^-}$ in the Case 1 of Equation~(\ref{cimp-caz1})---\textbf{left}, and in the Case 2 of Equation (\ref{cimp-caz2})---\textbf{right}.}
\label{fig-conc-werner-case-pura}
\end{figure}

  $C(\ket{\psi (t)})= \sqrt{1-4\, \frac{\tanh^2(\tau_-)}{\cosh^2(\tau_-)}\, \sin^2[\sinh(\tau_-)]}$   for the Case 2 of Equation (\ref{cimp-caz2}).
Both expressions were first written in Ref.~\cite{Messina-2016}.
The dynamics of the concurrence is shown in Figure~\ref{fig-conc-werner-case-pura}.
By using the analytical expression of concurrence of the state $\ket{\psi (t)}$, one obtains that the concurrence never vanishes in both cases of the applied fields.
If $\rho $ is the density operator of a two-qubit system, then its spin-flipped state is given by $\rho '= (\sigma_2\otimes \sigma_2)\, \rho^*\, (\sigma_2\otimes \sigma_2)$, where $\rho^*$ is the complex conjugate of $\rho $. The~matrix $\rho \, \rho '$ is a non-Hermitian matrix~\cite{Wootters-1998, Wootters-2001}, and it can be proven~\cite{Fan-2019} that its four eigenvalues are real and non-negative.
Let us denote these eigenvalues by $\nu _1$, $\nu _2$, $\nu _3$, and $\nu _4$, in decreasing order. The~concurrence is defined by
$
C(\rho)=\max\{\sqrt{\nu_1}-\sqrt{\nu_2}-\sqrt{\nu_3}-\sqrt{\nu_4},0 \}.
$

The expression of the concurrence of the Werner state (\ref{st-werner}) is~\cite{Ali}:
\[
C(\rho_W )=\max \left\{ \frac{3\, \alpha -1}{2}, 0 \right\}.
\]

For $\alpha \in (\frac{1}{3},1]$ the concurrence is greater than zero, which means that the Werner state (\ref{st-werner}) is~inseparable.

The expression of the concurrence of an arbitrary $X$ state was found in Ref.~\cite{Eberly-2007}:
\begin{equation}
C(\rho_{\mbox x})=2\, \max \left\{ 0, |\rho_{23}|-\sqrt{\rho_{11}\, \rho_{44}}, \, |\rho_{14}|-\sqrt{\rho_{22}\, \rho_{33}}\right\}.
\label{conc-X}
\end{equation}

Let us define the state $\ket{\xi} $ of two qubits as follows:
\begin{equation}
\ket{\xi}=\mu \ket{01}+\nu \ket{10},
\label{st-xi}
\end{equation}
with $\mu $ and $\nu $ complex parameters satisfying $|\mu|^2+|\nu|^2=1$.
We construct a special class of two-qubit mixed states, which includes the family of the Werner state, as follows:
\begin{equation}
\eta_{\mu ,\nu }^{(\alpha )}=\frac{1-\alpha }{4}\, I\otimes I +\alpha \, \proj{\xi}{\xi},
\label{st-gen-eta}
\end{equation}
where $\alpha \in [-\frac{1}{3},1]$ and the two complex parameters $\mu$ and $\nu$ satisfy the normalization condition of $\ket{\xi}$.  It is worth noticing that not all the states belonging to the class  defined by Equation (\ref{st-gen-eta}) are Werner–Popescu states, since some of them may be unitarily generated acting  independently on the two qubits. For $\mu =1/\sqrt 2$ and $\nu =-1/\sqrt 2$, the state $\eta_{\mu ,\nu }^{(\alpha )}$ becomes the Werner state (\ref{st-werner}). The~mixed states $\rho (t)$ of Equation (\ref{ro-t-st-fi}) is a subclass of the set of states $\eta_{\mu ,\nu }^{(\alpha )}$, obtained for the particular case $\mu =c_{01}(t)$ and $\nu =c_{10}(t)$, with $c_{01}$ and $c_{10}$ given by Equations (\ref{c01}) and (\ref{c10}).
The state (\ref{st-gen-eta}) is an $X$ state described by the non-zero elements:
\begin{eqnarray}
\rho_{11}&=&\rho_{44}=\frac{1-\alpha }{4},\nonumber \\
\rho_{22}&=&\frac{1-\alpha }{4}+\alpha\, |\mu |^2,\nonumber \\
\rho_{33}&=&\frac{1-\alpha }{4}+\alpha\, |\nu |^2,\nonumber \\
\rho_{23}&=&\alpha \, \mu \, \nu ^*. \nonumber
\end{eqnarray}

By using the expression of the concurrence of an $X$ state given by Equation (\ref{conc-X}), one obtains:
\begin{equation}
C(\eta_{\mu ,\nu }^{(\alpha )})
= \max \left\{ 0,  g\, (\alpha ,\mu )\right\},
\label{conc-eta-gen}
\end{equation}
where
\begin{equation}
g\, (\alpha ,\mu )=2\, |\alpha |\, |\mu |\, \sqrt{1-|\mu |^2}-\frac{1-\alpha }{2}.
\label{def-fct-g}
\end{equation}

Let us analyze in detail the expression of the concurrence, by investigating the two possible intervals for $\alpha $.
For $\alpha \in [-\frac{1}{3}, \frac{1}{3}]$, one has
$
g\, (\alpha ,\mu ) \le 0
$
for any $|\mu |\in [0,1]$ and, therefore, the concurrence is equal to zero:
\begin{equation}
C(\eta_{\mu ,\nu }^{(\alpha )})=0 \; \; \mbox{for}\; \mbox{any} \; |\mu |\in [0,1] \; \mbox{and} \; \alpha \in \left[-\frac{1}{3}, \frac{1}{3}\right].
\label{conc-zero-alfa-mic-o-treime}
\end{equation}

\begin{figure}
\centering
\includegraphics[width=7.5cm]{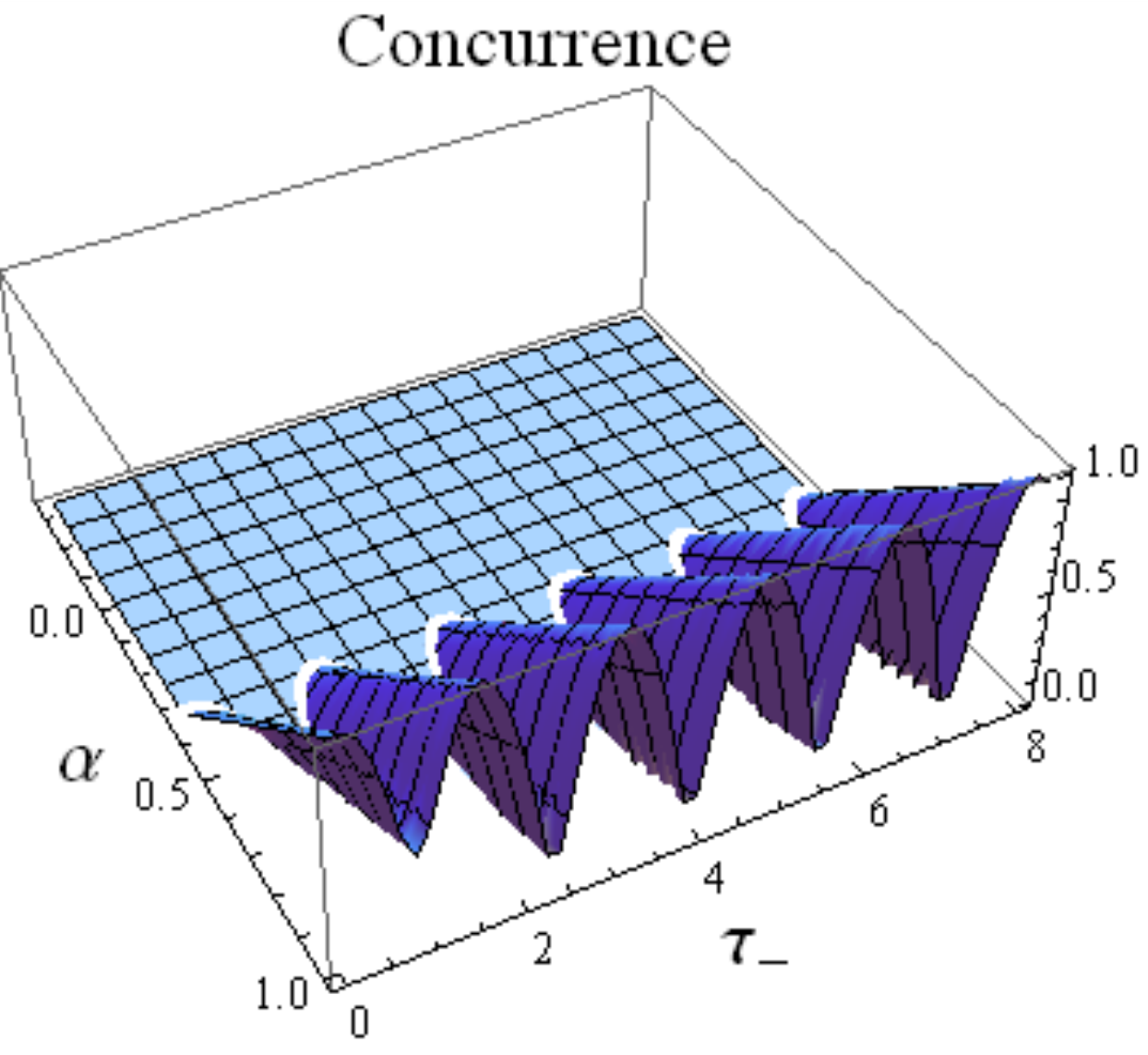}
\includegraphics[width=7.5cm]{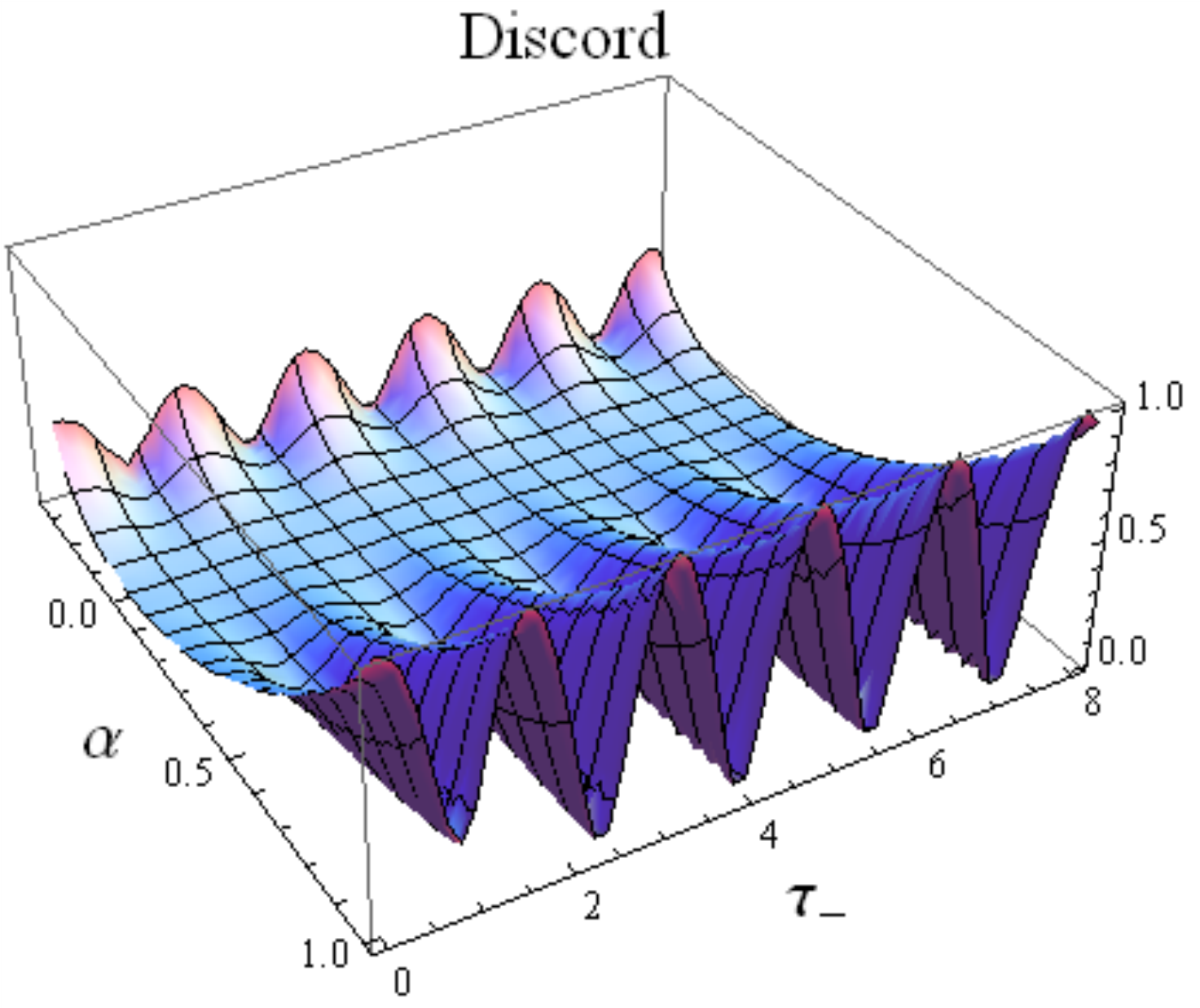}
\caption{Concurrence (left) and quantum discord (right) for the two-qubit system when the initial state is the Werner state (\ref{st-werner}) in the Case 1 of Equation (\ref{cimp-caz1}) in terms of the parameter $\alpha \in [-\frac{1}{3},1]$ of the Werner state and $\tau_-={|\Gamma_-| \over \hbar}\, t $ .}
\label{fig-werner-case1}
\end{figure}

\begin{figure}
\centering
\includegraphics[width=7.5cm]{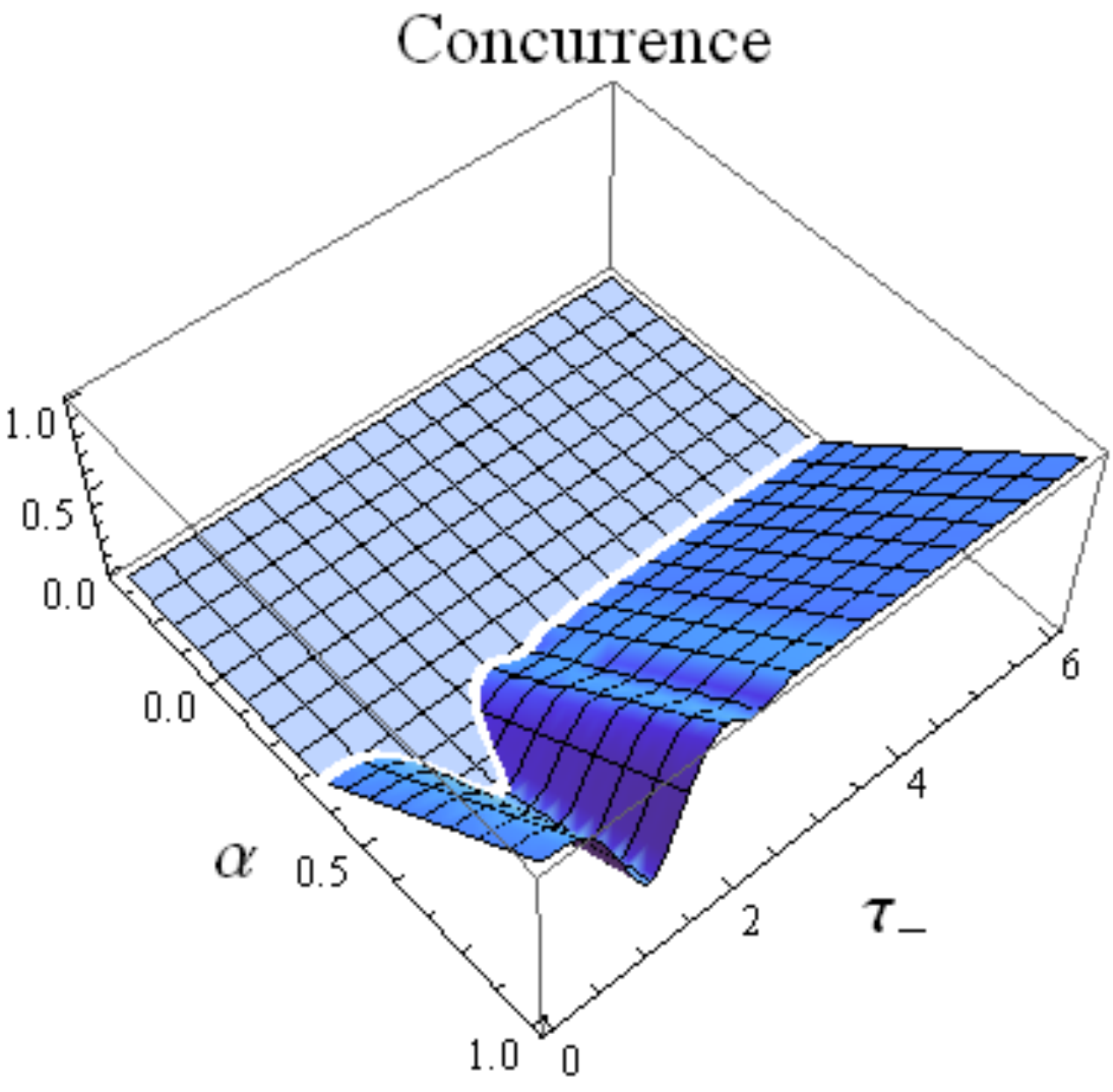}
\includegraphics[width=7.5cm]{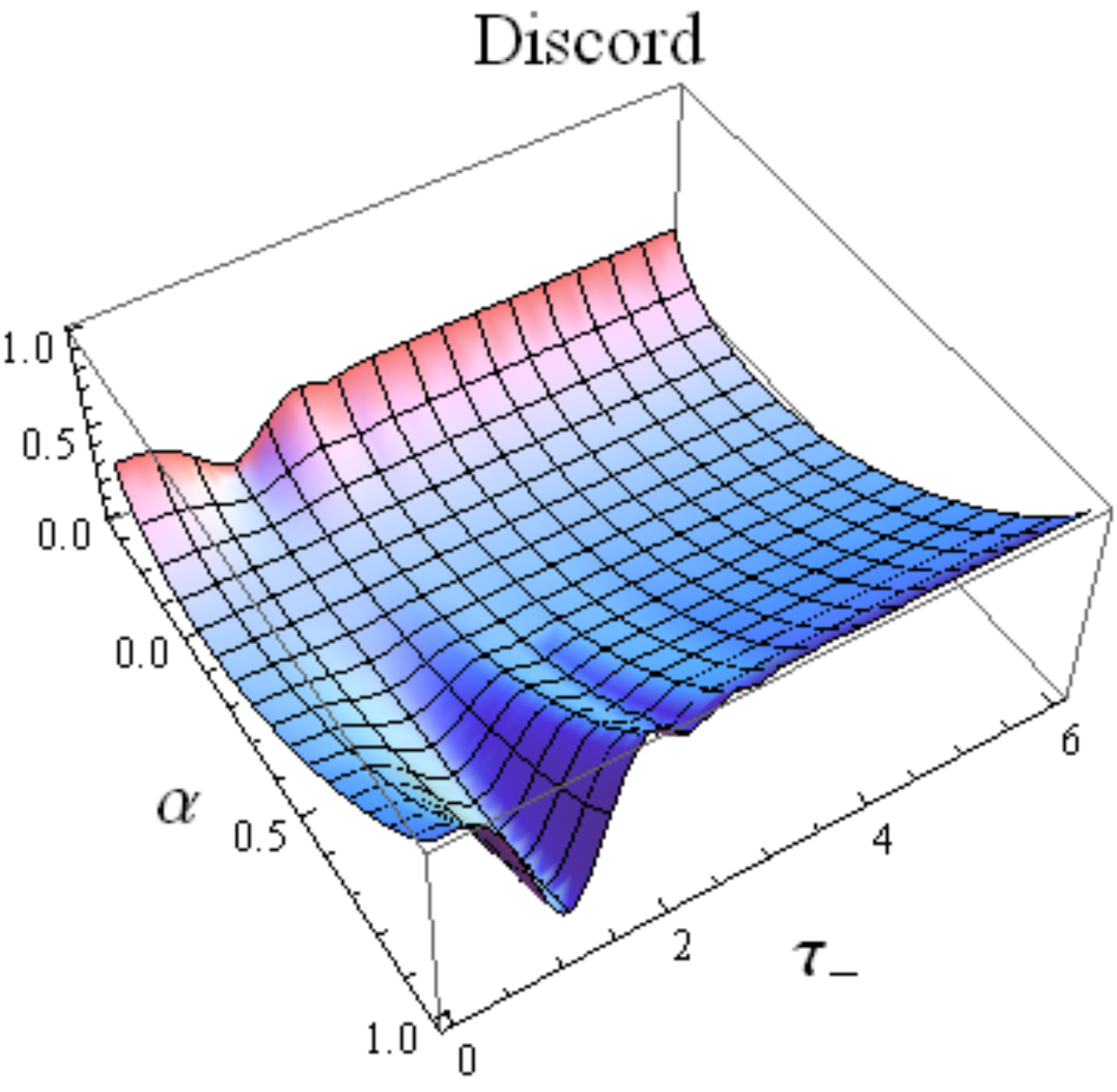}
\caption{Concurrence (left) and quantum discord (right) for the two-qubit system when the initial state is the Werner state (\ref{st-werner}) in the Case 2 of Equation (\ref{cimp-caz2})  in terms of the parameter $\alpha \in [-\frac{1}{3},1]$ of the Werner state and $\tau_-={|\Gamma_-| \over \hbar}\, t $.}
\label{fig-werner-case2}
\end{figure}

Since $\rho (t)$ is a subclass of the mixed states $\eta_{\mu ,\nu }^{(\alpha )}$, this fact explains the vanishing concurrence for $\rho(t)$ characterized by $\alpha \le 1/3$ for both cases of the two magnetic fields shown in Figures~\ref{fig-werner-case1} and~\ref{fig-werner-case2}-left.

For $\alpha \in \left( \frac{1}{3}, 1\right]$, instead, the equation $g\, (\alpha ,\mu )=0$ may be cast in the following form:
\begin{equation}
\alpha =\frac{1}{1+4\, |\mu |\, \sqrt{1-|\mu |^2}}.
\label{expr-alfa-zero}
\end{equation}

\begin{figure}
\centering
\includegraphics[width=10cm]{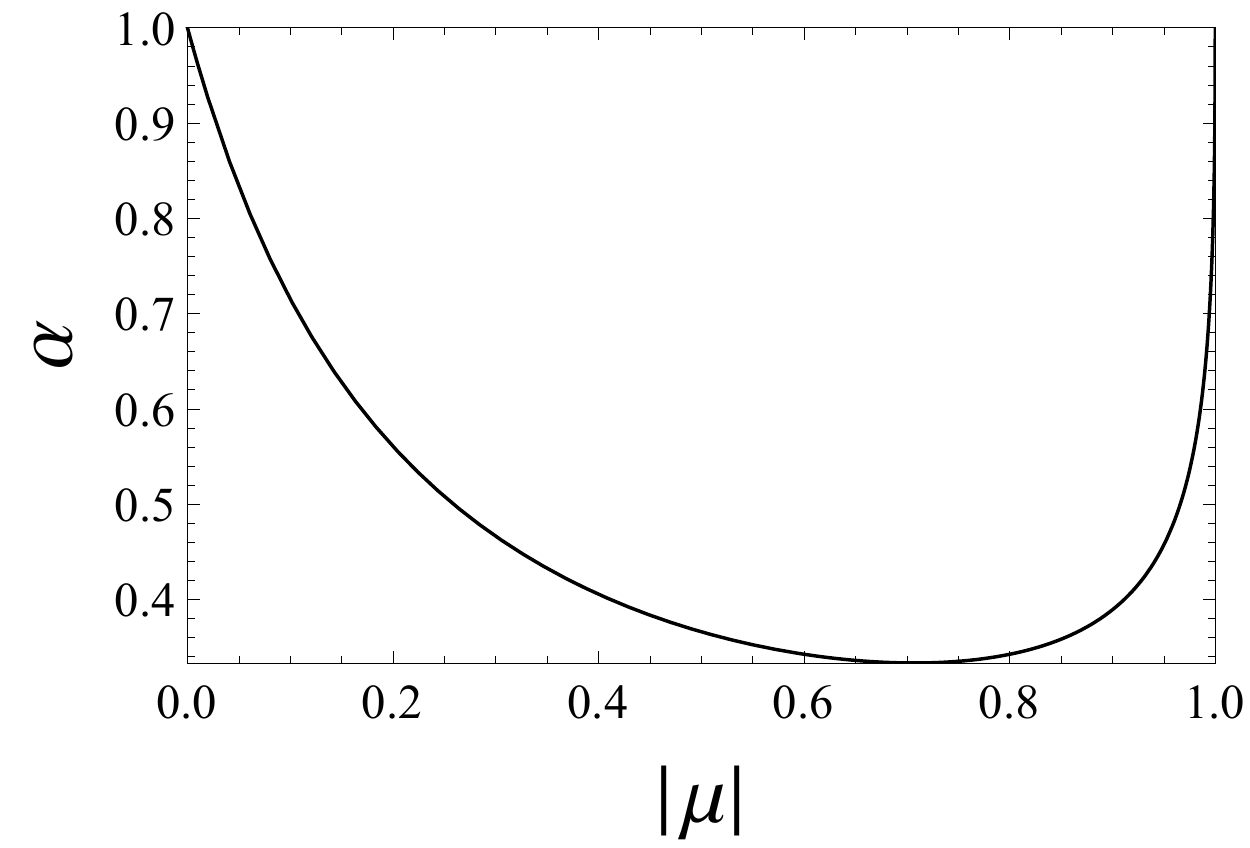}
\caption{Plot of $\alpha $ in terms of $|\mu |$ by using Equation (\ref{expr-alfa-zero}) for which the generalized Werner state $\eta_{\mu ,\nu }^{(\alpha )} $ is characterized by a vanishing concurrence.}
\label{fig-alfa-conc-zero}
\end{figure}

If we represent Equation (\ref{expr-alfa-zero}) in the $\alpha$-$|\mu|$ plane, the curve $\alpha(|\mu|)$ distinguishes the region wherein the concurrence vanishes from the one where the concurrence is positive.
In other words, Equation (\ref{expr-alfa-zero}) defines in the $\alpha$-$|\mu|$ plane the border between appearance and disappearance of entanglement between the two spins within the class of the generalized Werner states $\eta_{\mu,\nu}^\alpha$.
In particular when $\alpha \leq 1/3$ the concurrence is zero whatever $\mu$ is.
When, instead, $\alpha > 1/3$ there always exists an $\alpha$-dependent interval $[|\mu_1|,|\mu_2|]$ within which the concurrence is different from zero.
In Figure~\ref{fig-alfa-conc-zero} we plot $\alpha $ in terms of $|\mu |$ by using Equation (\ref{expr-alfa-zero}) for which the concurrence of the state $\eta_{\mu ,\nu }^{(\alpha )} $ is equal to zero.

We obtain the following expression of the concurrence of the state $\eta_{\mu ,\nu }^{(\alpha )} $:
\[
C(\eta_{\mu ,\nu }^{(\alpha )})=\left\{ \begin{array}{ccl}
0 & \mbox{for } & |\mu |\in  \left[ 0, \frac{1}{2} -\frac{\sqrt{3\alpha^2+2\alpha -1}}{4 \alpha}  \right] \\
2\, \alpha \, |\mu |\, \sqrt{1-|\mu |^2}-\frac{1-\alpha }{2}  & \mbox{for } & |\mu | \in \left( \frac{1}{2} -\frac{\sqrt{3\alpha^2+2\alpha -1}}{4 \alpha} , \frac{1}{2} +\frac{\sqrt{3\alpha^2+2\alpha -1}}{4 \alpha}  \right) \\
0 & \mbox{for } & |\mu |\in  \left[ \frac{1}{2} +\frac{\sqrt{3\alpha^2+2\alpha -1}}{4 \alpha}, 1  \right]
\end{array} \right.
\]

It is easy to see that $C(\eta_{\mu ,\nu }^{(\alpha )})=C(\rho_W)=(3\alpha-1)/2$ under the condition $|\mu|=1/\sqrt{2}$.
This implies, in particular, that we get the same value of the concurrence of $\rho_W$ [Equation (\ref{st-werner})] if we substitute $\ket{\Psi^-}$ with~$\ket{\Psi^+}$.

\begin{figure}
\centering
\includegraphics[width=10cm]{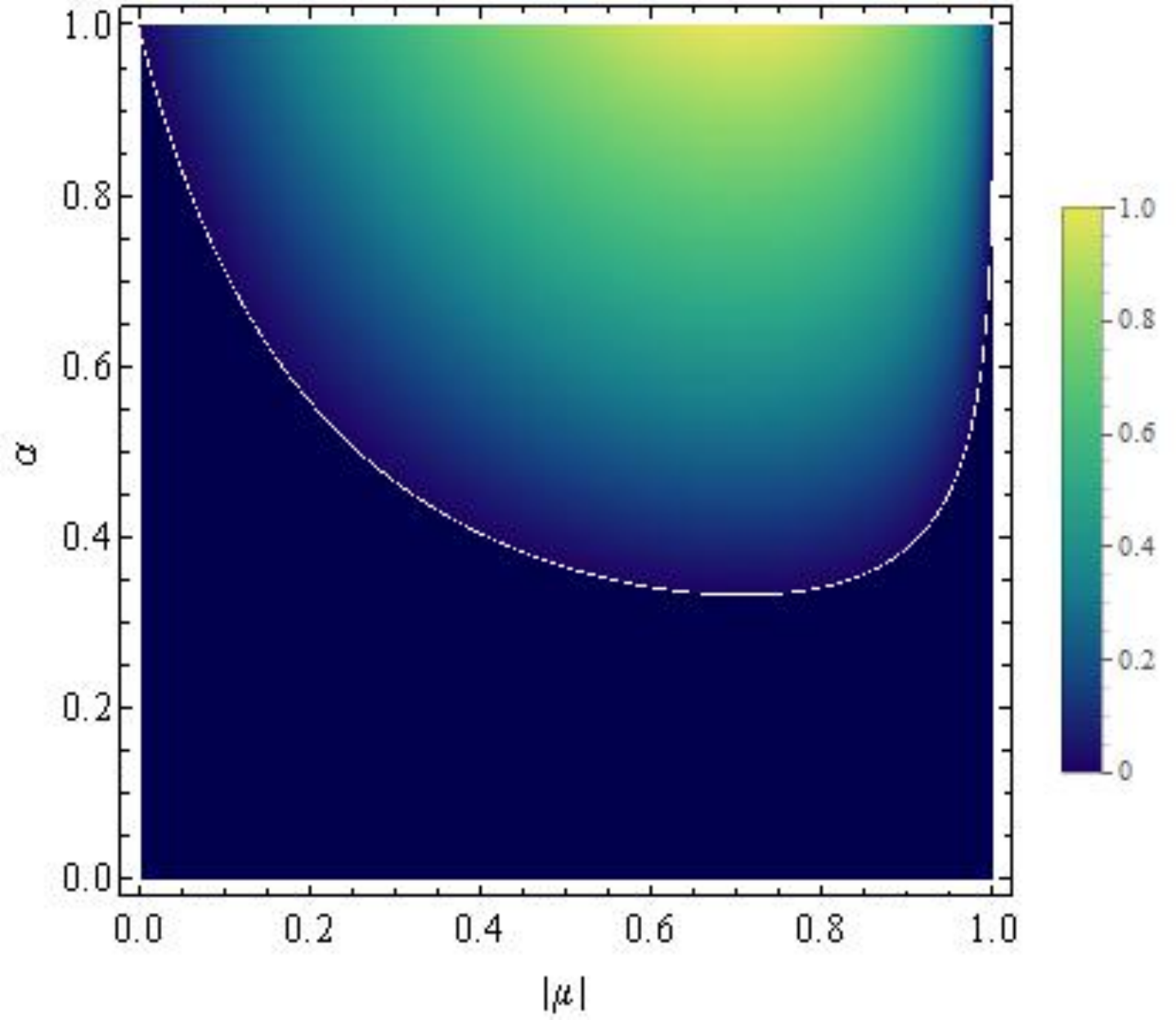}
\caption{Concurrence of the generalized Werner states $\eta_{\mu,\nu}^{(\alpha )}$ in terms of $\alpha $ and $|\mu |$. }
\label{fig-conc-density-plot}
\end{figure}

It is worth noticing, in addition, that
\[
\eta_{\mu ,\nu }^{(\alpha )}(t)=\frac{1-\alpha }{4}\, I\otimes I +\alpha \, \proj{\xi(t)}{\xi(t)},
\]
meaning that the generalized Werner states $\eta_{\mu,\nu}^{(\alpha )}$ evolve keeping their $\alpha$-dependent structure.
Hence the time evolution of a generalized Werner state characterized by a particular value of $\alpha$ generates only ``horizontal movements'' in the $\alpha$-$|\mu|$ plane in Figure~\ref{fig-conc-density-plot}.
This circumstance implies that during its time evolution, a generalized Werner state may enter into or go out the non-zero-concurrence region identified in Figure~\ref{fig-conc-density-plot}.
For example, if we consider the entangled generalized Werner state defined by $\alpha=\mu=0.5$ as the initial condition, it may happen that at a certain time instant, $\mu$ becomes less than $\approx$0.25.
In this case, then, a sudden death of entanglement is exhibited.
Of course, if $|\mu|$ comes back to its original value in a finite interval of time, a rebirth of entanglement would follow a plateau of zero concurrence.
Such a possibility is confirmed by the plots reported in a following subsection, where we compare the concurrence and the quantum discord in time for our two-spin system under the two exactly solvable time-dependent scenarios (\ref{cimp-caz1}) and (\ref{cimp-caz2}).

Suppose that the initial state of the two qubits is the Werner state $\rho_W^{(\alpha )}$. By using Equation (\ref{conc-eta-gen}), we find the analytical expression of the concurrence of the evolved Werner state $\rho (t)$ of Equation (\ref{ro-t-st-fi}). When the applied magnetic fields have the expression of Equation (\ref{cimp-caz1}), i.e., Case 1, we get:
\begin{equation}
C(\rho (t))=\max \left\{ 0, |\alpha | \sqrt{1-\tanh ^2(2\tau_-)\sin^2(2\tau_-)}-\frac{1-\alpha }{2} \right\}.
\label{conc-caz-1}
\end{equation}

The analytical expression of the concurrence when the magnetic fields are described by Case 2, i.e., by Equation (\ref{cimp-caz2}), is given by:
\begin{equation}
C(\rho (t))=\max \left\{ 0, |\alpha | \sqrt{1-4\, \frac{\tanh^2(\tau_-)}{\cosh^2(\tau_-)}\, \sin^2[\sinh(\tau_-)]}   -\frac{1-\alpha }{2} \right\}.
\label{conc-caz-2}
\end{equation}

For obtaining the analytical expressions of the concurrence given by Equations (\ref{conc-caz-1}) and (\ref{conc-caz-2}), we have employed Equations (\ref{a4}), (\ref{a6}) and (\ref{a7}) for Case 1, and Equations (\ref{a9}), (\ref{a11}) and (\ref{a12}) for Case 2, respectively. In addition, we have considered $\gamma _{12}=\gamma_{21}$, which leads to $\phi_{\Gamma_-}=0$ (see Equation (\ref{fi-gama})).

We plot the concurrence of the state $\rho (t)$ in terms of the parameter $\alpha$ of the initial Werner state and $\tau_-$ defined by Equation (\ref{tau-pm}) in Figures~\ref{fig-werner-case1} and~\ref{fig-werner-case2}-left. From them one can notice that the concurrence is equal to zero for $\rho(t)$ characterized by $\alpha \le 1/3$ for both cases, Case 1 and Case 2, of the two applied magnetic fields. We have presented an analytical proof of this fact, by showing that according to Equation (\ref{conc-zero-alfa-mic-o-treime}), zero concurrence occurs for the generalized Werner state $\eta_{\mu , \nu }^{(\alpha )} $ for the particular $\alpha $ satisfying $\alpha \le 1/3$ for any value of $\mu $.

\subsection{Quantum Discord}
\label{sec-disc}

A different important measure of quantum correlations we investigate in this paper is quantum discord. The~quantum discord can be evaluated for an $X$ state by using the approach presented in Appendix~\ref{sec-discord}. Since the evolved Werner state $\rho (t)$ is an $X$-state, we can use the results given in Appendix~\ref{sec-discord} for computing the quantum discord according to Equation (\ref{discord}):
$D(\rho_{AB})={\cal I}(\rho_{AB})-{\cal C}(\rho_{AB}). $
We plot, in addition, quantum discord of $\rho (t)$ in terms of $\alpha$ and $\tau_-$ in Figures~\ref{fig-werner-case1} and~\ref{fig-werner-case2}-right.

\subsection{Comparison between the Concurrence and Quantum Discord of the Evolved Werner State}
\label{sec-comp}

Our purpose in this subsection is to make a detailed comparison between the concurrence and quantum discord of a given evolved Werner state $\rho (t)$, i.e., for a fixed value of $\alpha $.

\begin{figure}
\centering
\includegraphics[width=7.5cm]{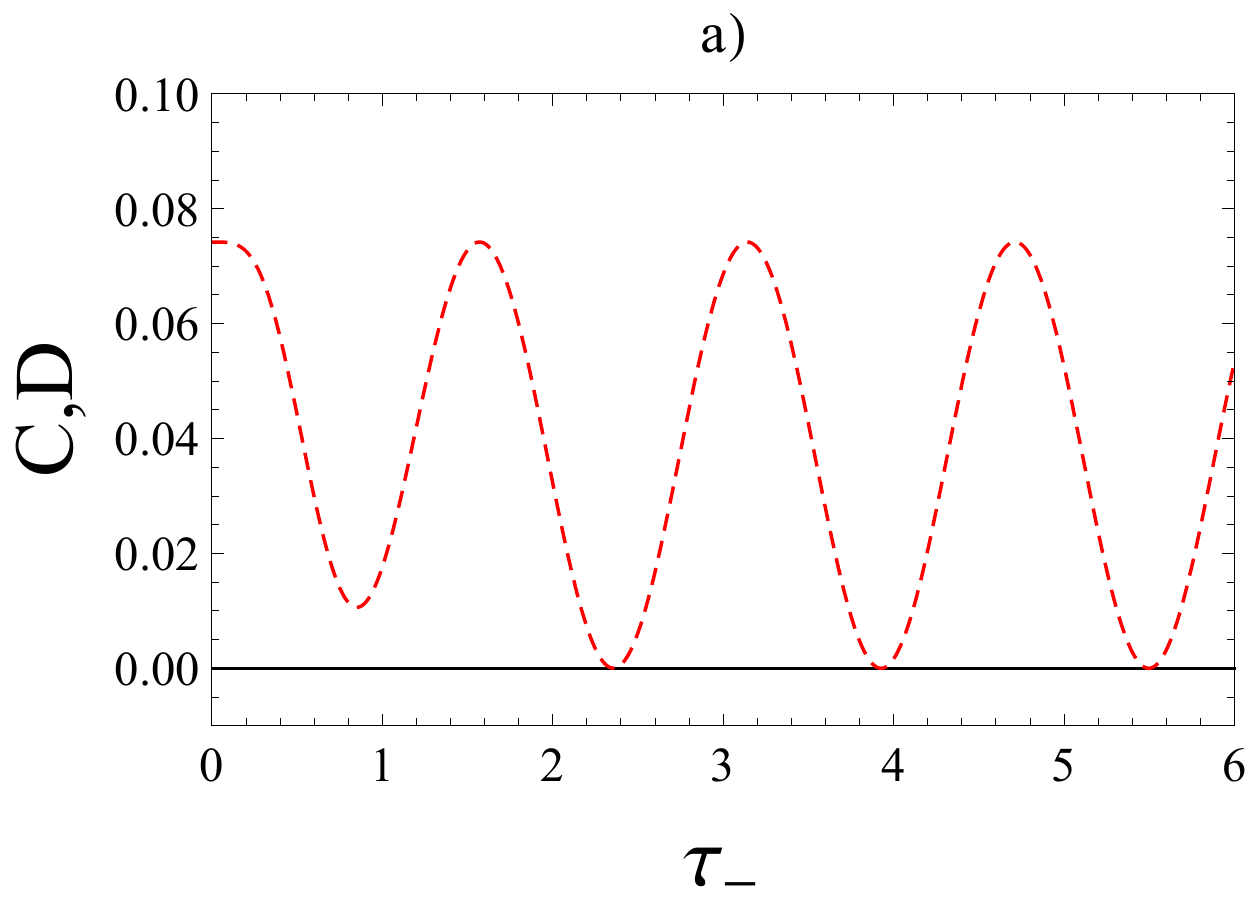}
\includegraphics[width=7.5cm]{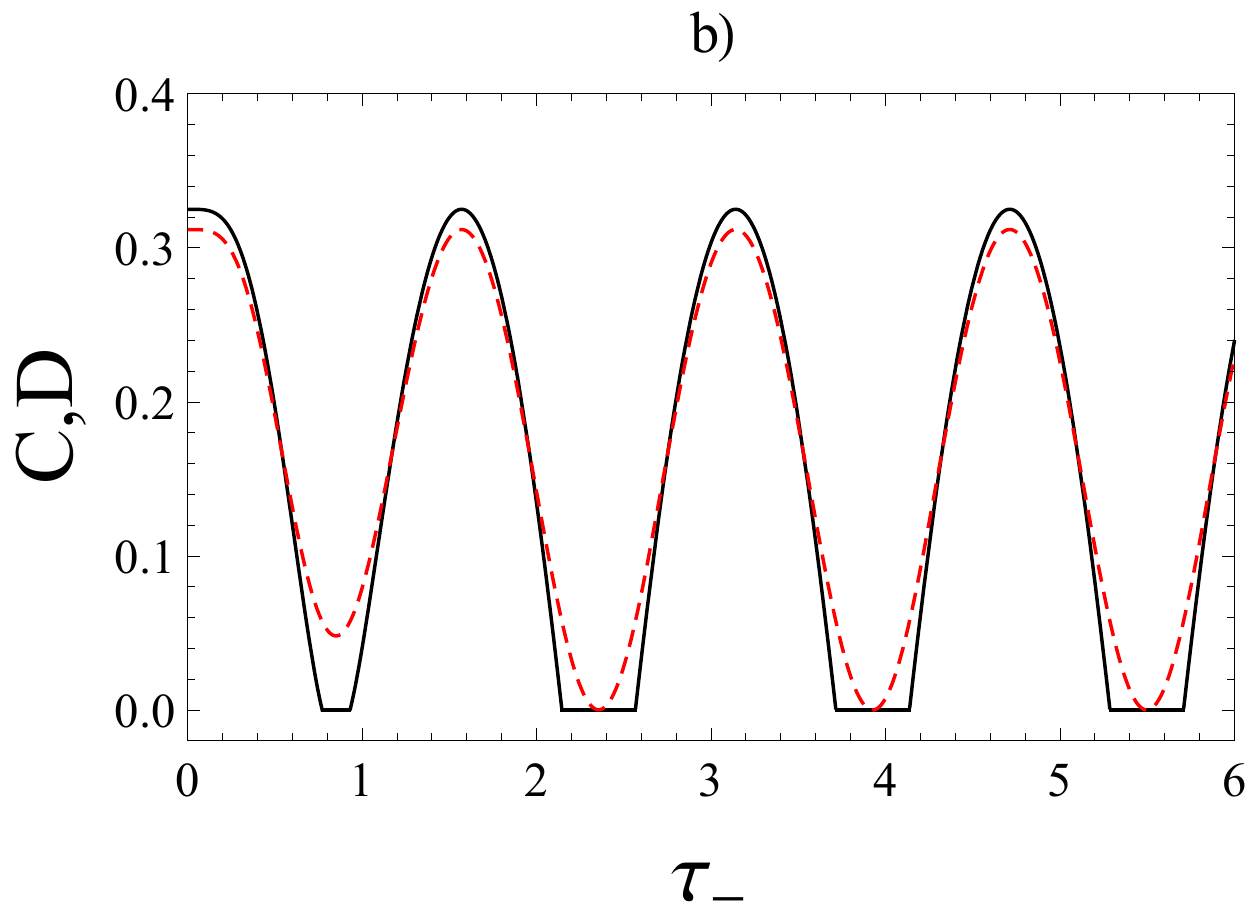}
\caption{Concurrence (black, solid) and quantum discord (red, dashed) when the state at $t=0$ is the Werner state (\ref{st-werner}) in the Case 1 of Equation (\ref{cimp-caz1}) in terms of  $\tau_-={|\Gamma_-| \over \hbar}\, t $ for: (\textbf{a}) $\alpha =0.25 $, (\textbf{b}) $\alpha =0.55$.}
\label{fig-werner-case1-plan}
\end{figure}

\begin{figure}
\centering
\includegraphics[width=7.5cm]{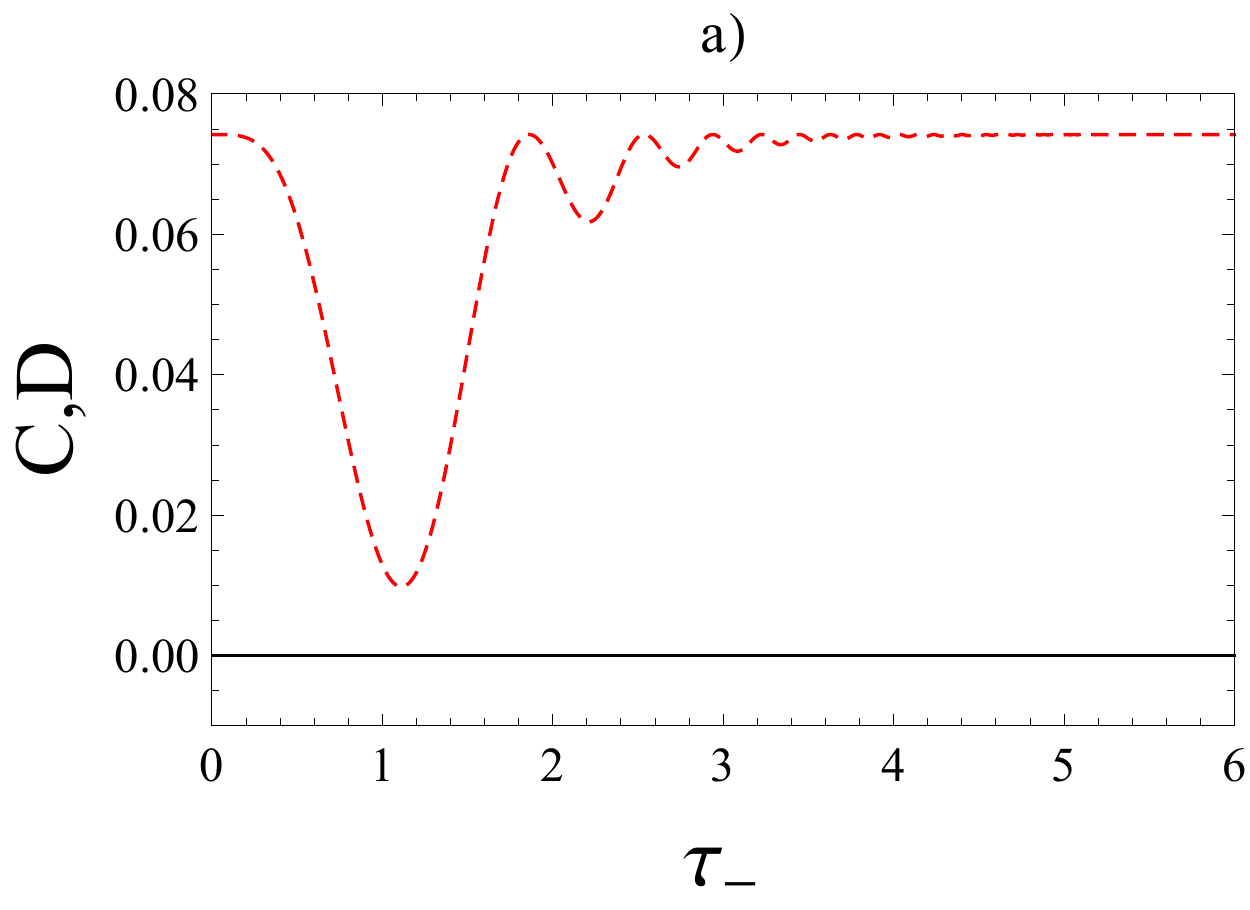}
\includegraphics[width=7.5cm]{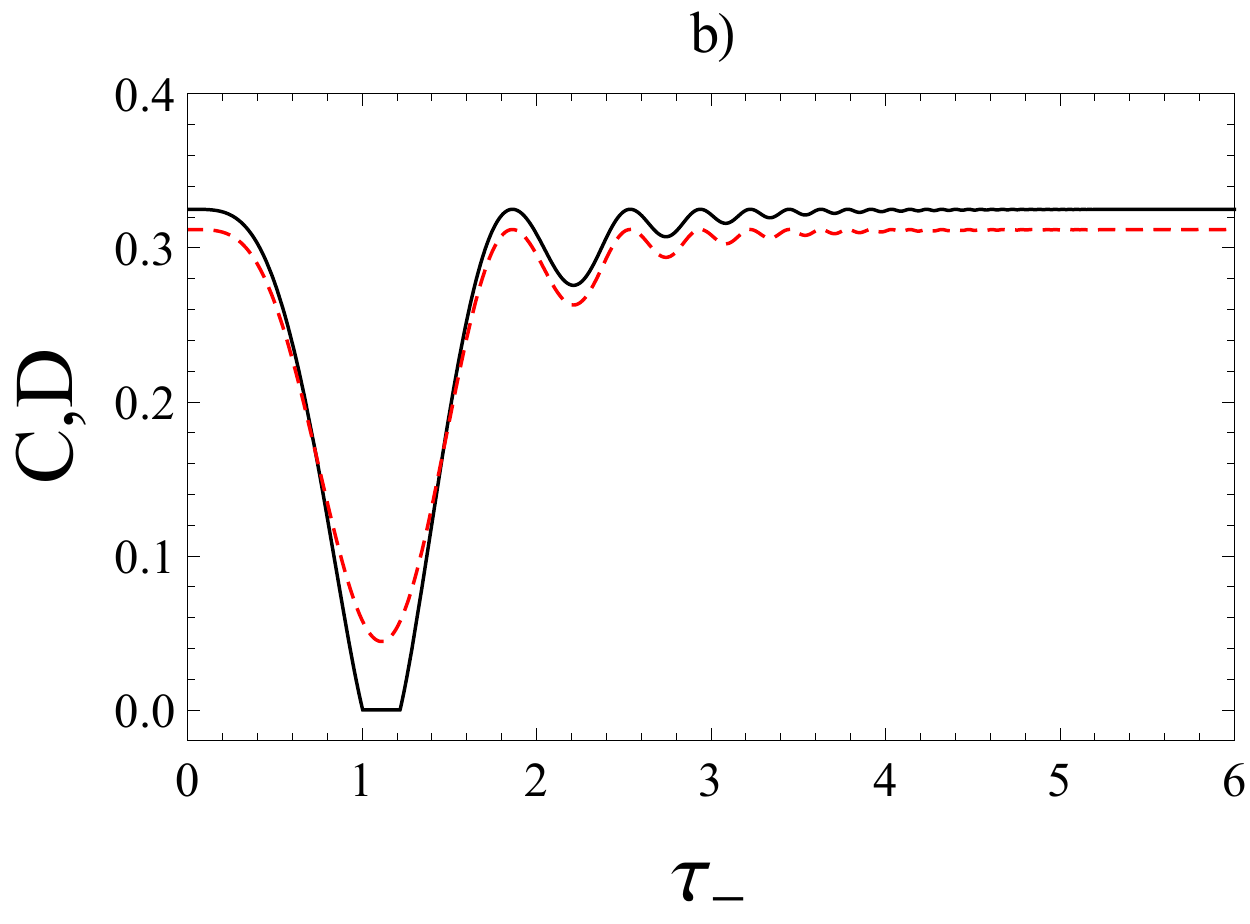}
\includegraphics[width=7.5cm]{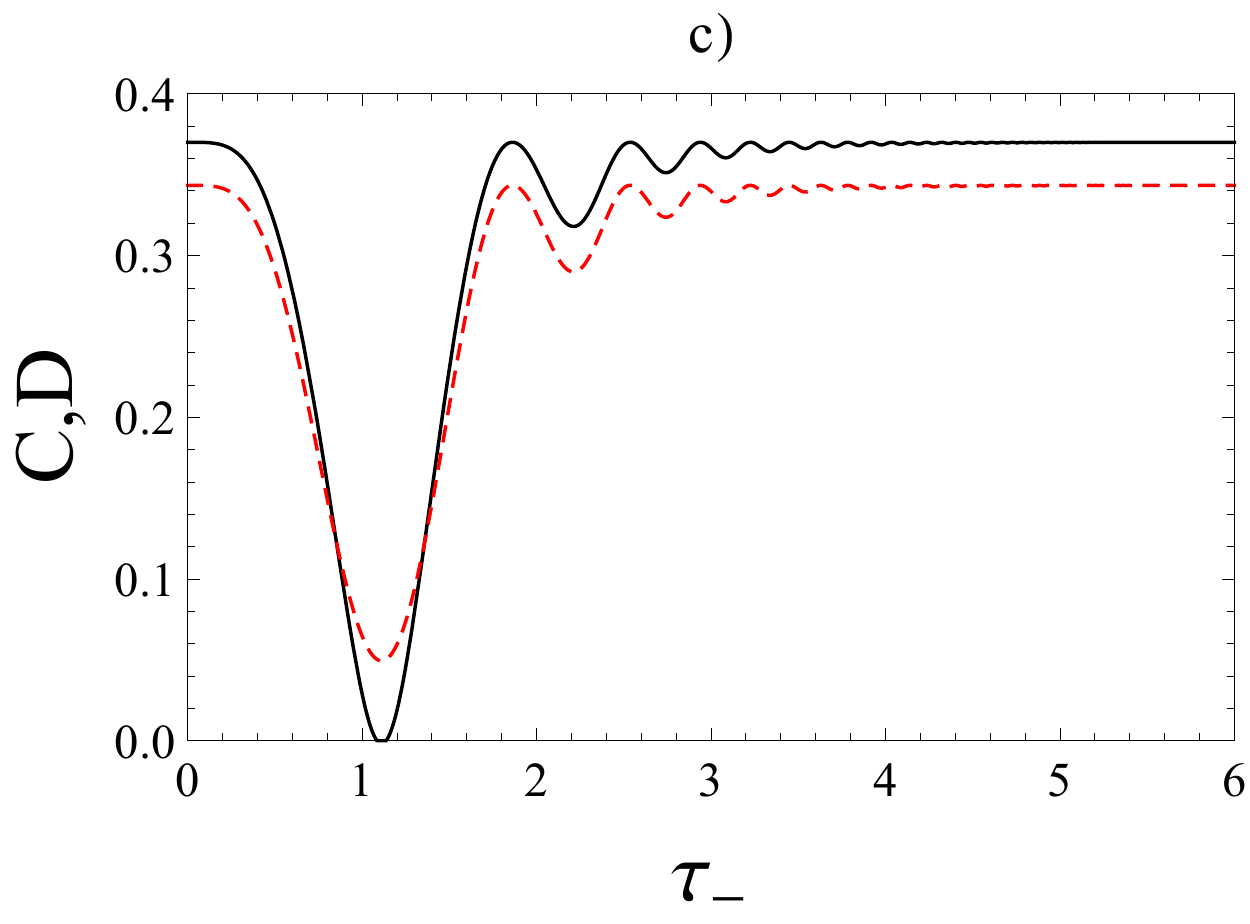}
\includegraphics[width=7.5cm]{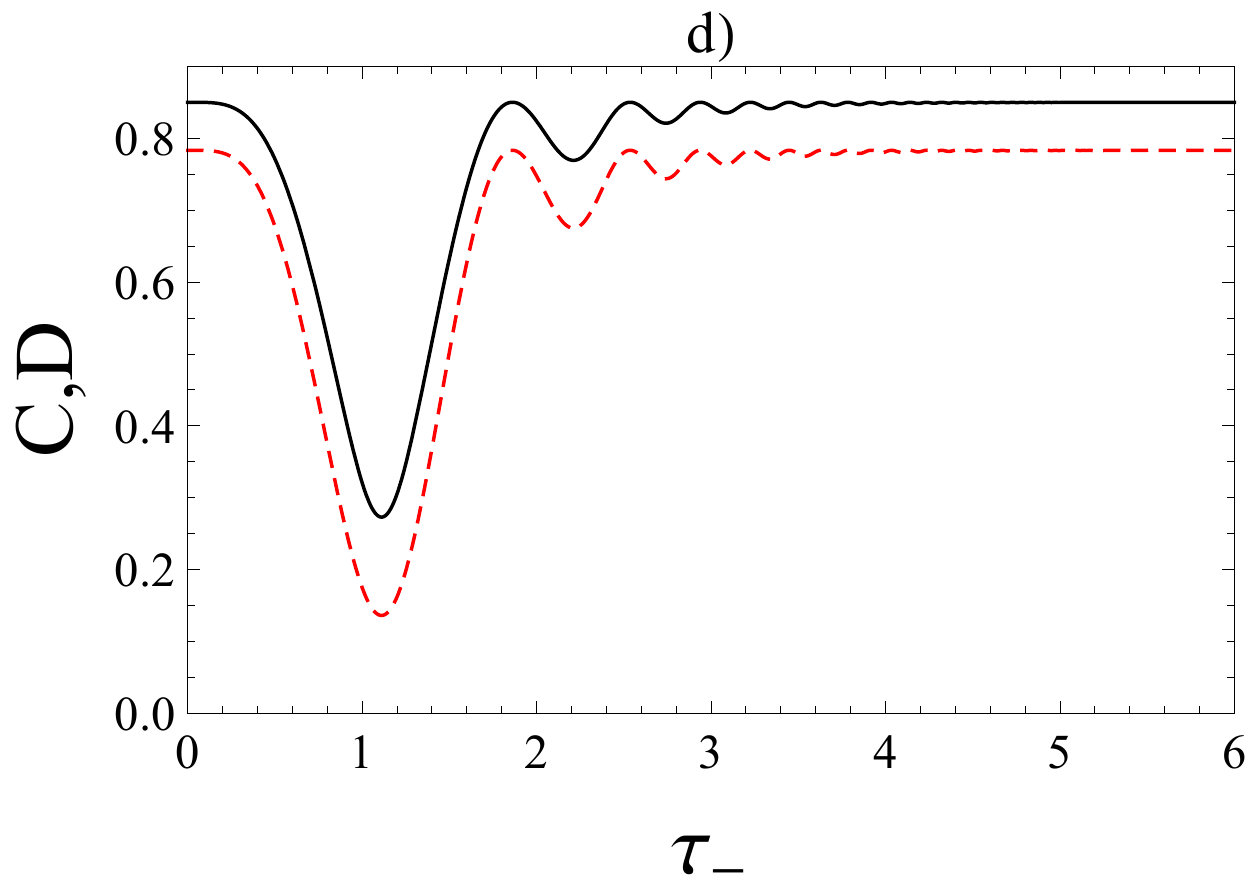}
\caption{Concurrence (black, solid) and quantum discord (red, dashed) when the state at $t=0$ is the Werner state (\ref{st-werner}) in the Case 2 of Equation (\ref{cimp-caz2})  in terms of $\tau_-={|\Gamma_-| \over \hbar}\, t $ for: (\textbf{a}) $\alpha =0.25 $, (\textbf{b}) $\alpha =0.55$, (\textbf{c})~$\alpha = 0.582$, (\textbf{d}) $\alpha =0.9 $.}
\label{fig-werner-case2-plan}
\end{figure}

We present the evolution of both concurrence and quantum discord in terms of $\tau_-$ defined by Equation (\ref{tau-pm}) for Case 1 and Case 2 of the applied magnetic fields.
For $\alpha \in \left[ -\frac{1}{3}, \frac{1}{3}\right]$, both in Case 1 and Case 2, the concurrence is equal to zero as expected (see Figure~\ref{fig-werner-case1-plan}a and Figure~\ref{fig-werner-case2-plan}a).
For $\alpha \in \left(\frac{1}{3}, 1\right)$ in Case 1, there are zero-plateaux for concurrence and the discord is non-zero (see Figure~\ref{fig-werner-case1-plan}b). In this case, the phenomenon of sudden death of entanglement, followed by revival of entanglement occur many times.

In Case 2, for $\alpha \in \left(\frac{1}{3}, 0.582\right)$, there is a unique zero-plateau for concurrence and the discord is non-zero {[see Figure~\ref{fig-werner-case2-plan}b]}.
It is interesting to note that such a zero-plateau zone reduces to a single point when $\alpha \approx 0.582$ at the time instant $\tau_- \approx 1.115$ (see Figure~\ref{fig-werner-case2-plan}c). The~quantum discord, instead, remains different from zero: $D \approx 0.049 $.
For $\alpha \in \left( 0.582, 1\right)$, the phenomenon of sudden death and revival disappears since in this case the concurrence, as well as the quantum discord, is larger than zero (see Figure~\ref{fig-werner-case2-plan}d).

We emphasize that the plots reported and discussed above confirm the predictions exposed in a previous subsection.
Such plateaux, indeed, can be explained in the light of the observation based on the $\alpha-|\mu|$ relation in Equation (\ref{expr-alfa-zero}).

Finally, we underline that in Ref.~\cite{Xia-2018}, Xia~et~al. have found an analogue process to our case~(c) above on concurrence, but for quantum discord. They have investigated the dynamics of an open system, where the quantum channel was a stochastic dephasing channel along the $z$-direction. In Figure 7 of Ref.~\cite{Xia-2018}, they have shown that sudden death and sudden birth of quantum discord occur for a two-qubit Bell-diagonal state.
which presents a curve with the minimum value zero for quantum discord. One knows that if the quantum discord is equal to zero, then the concurrence is also zero, since a zero-discord state is separable. Therefore, in Ref.~\cite{Xia-2018}, they have presented sudden death and birth of both quantum discord and concurrence.

A further interesting investigation to be made is the case of constant magnetic fields applied on the two qubits found initially in the Werner state. This needs a different treatment as shown in Appendix~\ref{constant}. A new parameter $\beta $ is introduced, which influences the behavior of the shape of both concurrence and quantum discord as one can see in Figures~\ref{const-1} and~\ref{const-2}.

\section{Dynamical Origin of the Asymptotic Behavior of Quantum Correlations}
\label{sec-fid}

\begin{figure}
\centering
\includegraphics[width=7.5cm]{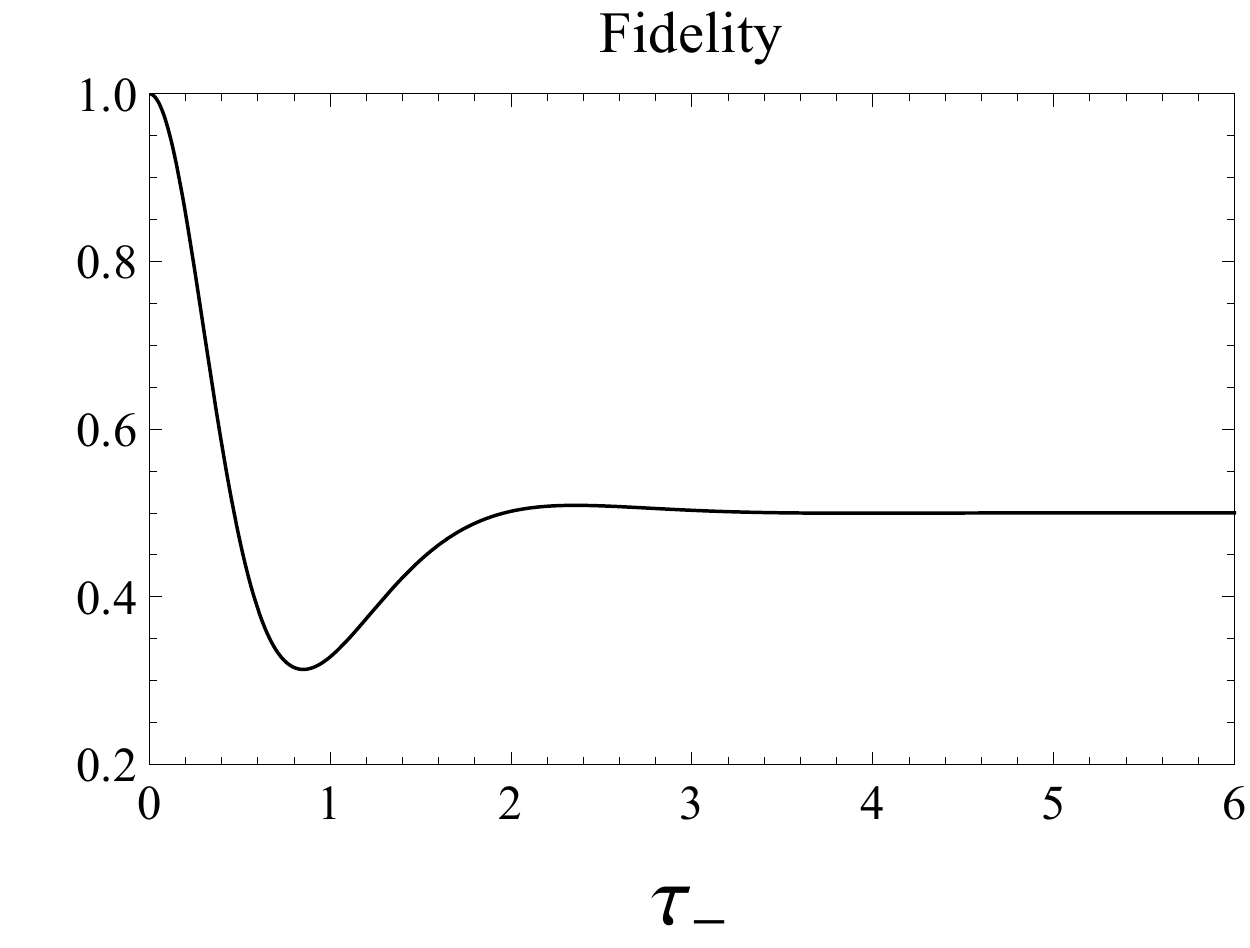}
\includegraphics[width=7.5cm]{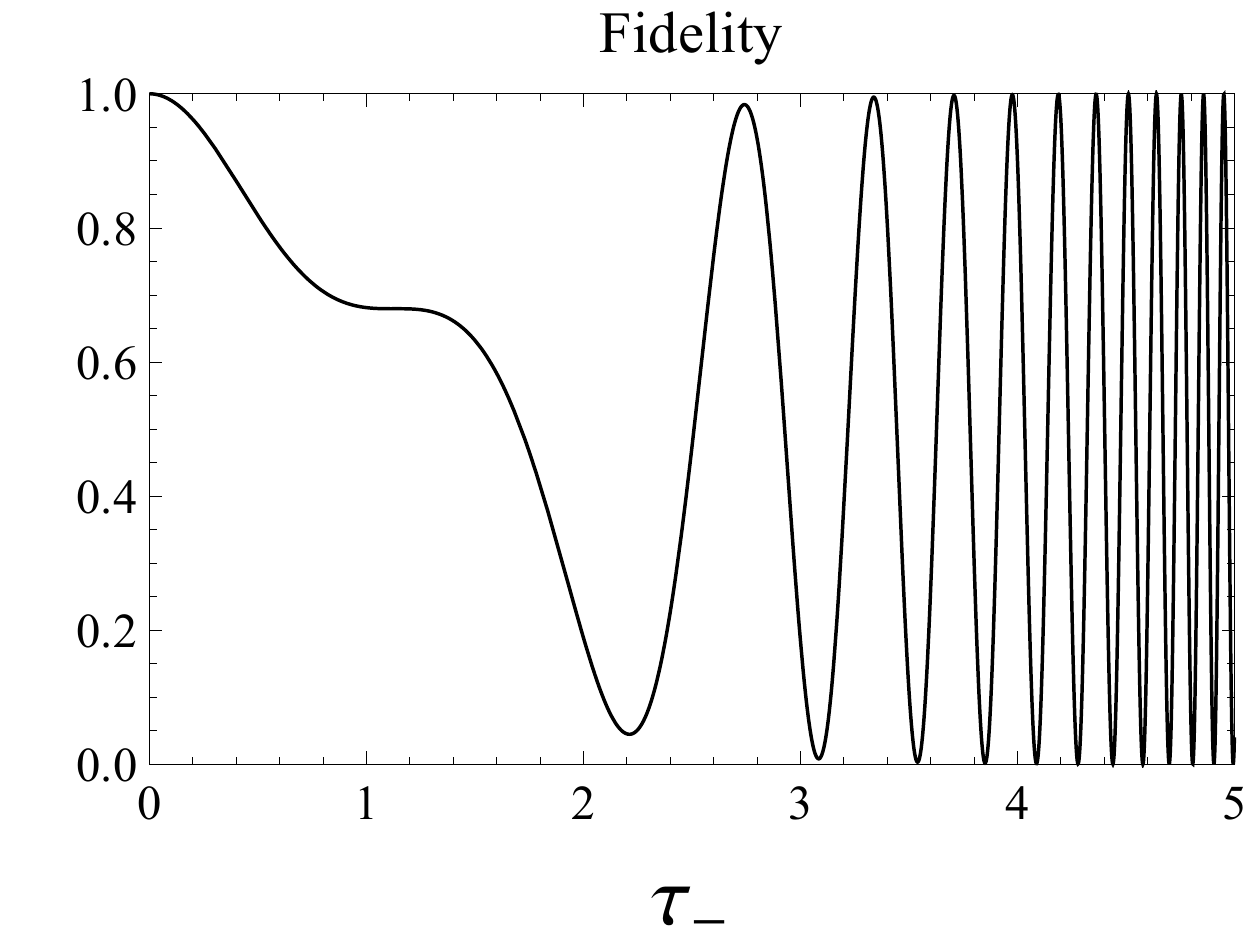}
\caption{Fidelity when the initial state is the singlet state, i.e., $\alpha =1$ in Case 1---\textbf{left} and in Case~2---\textbf{right}.}
\label{fig-fid-pura-werner-case}
\end{figure}

In this section, we provide a dynamical interpretation of the asymptotic behavior of both the concurrence and quantum discord exhibited by the system.
To this end we evaluate the time dependence of the fidelity of the Bell state $\ket{\Psi^-}$ with respect to its evolved state $\ket{\psi(t)}=U(t)\, \ket{\Psi^-}$, defined in Equation (\ref{psi-timp}), getting:
\begin{equation}
{\cal F}(\ket{\Psi^-},\ket{\psi(t)})=|\average{\psi(t)|\Psi^-}|^2= |a_-(t)|^2 \cos ^2\left( \phi_a^-\right) +|b_-(t)|^2 \sin ^2\left( \phi_b^-\right),
\label{fid-expr-pura}
\end{equation}
in accordance with Equations (\ref{c01}) and (\ref{c10}).
In Figure~\ref{fig-fid-pura-werner-case} we plot the fidelity (\ref{fid-expr-pura}) versus $\tau_-$ highlighting different asymptotic behaviors for large $\tau_-$ in the two cases, constant and oscillatory, respectively.
Exploiting Equations (\ref{a6}) and (\ref{a7}) one easily confirms that in Case 1 the asymptotic behavior of ${\cal F}(\ket{\Psi^-},\ket{\psi(t)})$ is time-independent and equal to 1/2.
This circumstance suggests that the asymptotic evolved state $\ket{\psi(\tau_- \gg 1)}$ is an equally weighted coherent superposition of the Bell states $\ket{\Psi^+}$ and $\ket{\Psi^-}$.
Such an intuitive prediction may be analytically supported mathematically acquiring the following form
\begin{equation}\label{Asymptotic State}
\ket{\psi(\tau_-\gg 1)} \approx {-ie^{-2i\tau_-}\ket{\Psi^+}+\ket{\Psi^-} \over \sqrt{2}}.
\end{equation}

The concurrence of this state reads $C=|\cos(2\tau_-)|$ and reproduces the asymptotic oscillations exhibited by the concurrence $C(\ket{\psi(\tau_-)})$ in Figure~\ref{fig-conc-werner-case-pura}-left.
The structure of $\ket{\psi(\tau_- \gg 1)}$, as given by Equation~(\ref{Asymptotic State}), transparently explains the dynamical origin of the oscillations dominating the time evolution of the concurrence (as well as of the quantum discord) for large $\tau_-$.

In view of Equations (\ref{a11}) and (\ref{a12}), the fidelity in Case 2, instead, asymptotically exhibits infinitely many maxima closer and closer to one as well as infinitely many minima closer and closer to zero.
Such a behavior is well illustrated in Figure~\ref{fig-fid-pura-werner-case}-right and suggests that the system asymptotically tends to reach a complete oscillatory regime between the states $\ket{\Psi^-}$ and $\ket{\Psi^+}$.
Even in this case such a prediction may be legitimated evaluating $\ket{\psi(\tau _-\gg 1)}$ related to Case 2, which can be cast in the following form
\begin{equation}\label{Asymptotic State 2}
\ket{\psi(\tau_-\gg 1)} \approx -\cos\left( {\sinh(\tau_-) \over 2} -{3\pi \over 4} \right)\ket{\Psi^+}-i\sin\left( {\sinh(\tau_-) \over 2} -{3\pi \over 4} \right)\ket{\Psi^-}.
\end{equation}

One can easily check that the concurrence for such a state reaches its maximum value $C=1$.
It is possible to interpret such a result claiming that the system goes from $\ket{\Psi^-}$ to $\ket{\Psi^+}$ and back through states whose concurrence is closer and closer to one as time goes on.
Incidentally, examining Equation~(\ref{Asymptotic State 2}), one can convince oneself that the semi-period of these oscillations progressively vanishes.
Thus, as in Case 1, the structure (\ref{Asymptotic State 2}) of $\ket{\psi(\tau_- \gg 1)}$ in Case 2 transparently brings to light the dynamical origin of the plateaux exhibited by the concurrence as well as by the quantum discord in Figure~\ref{fig-fid-pura-werner-case}-right.

\begin{figure}
\centering
\includegraphics[width=7.5cm]{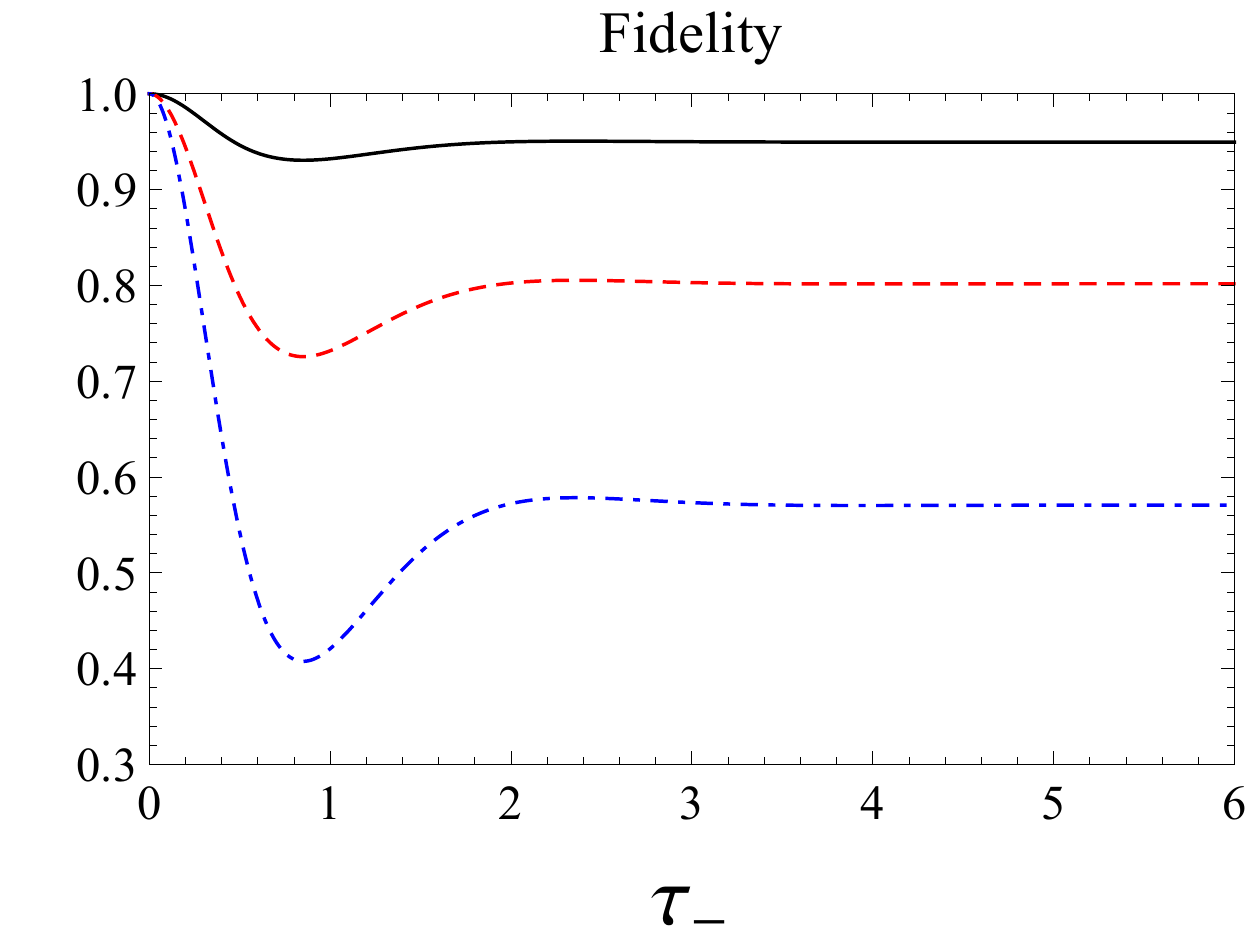}
\includegraphics[width=7.5cm]{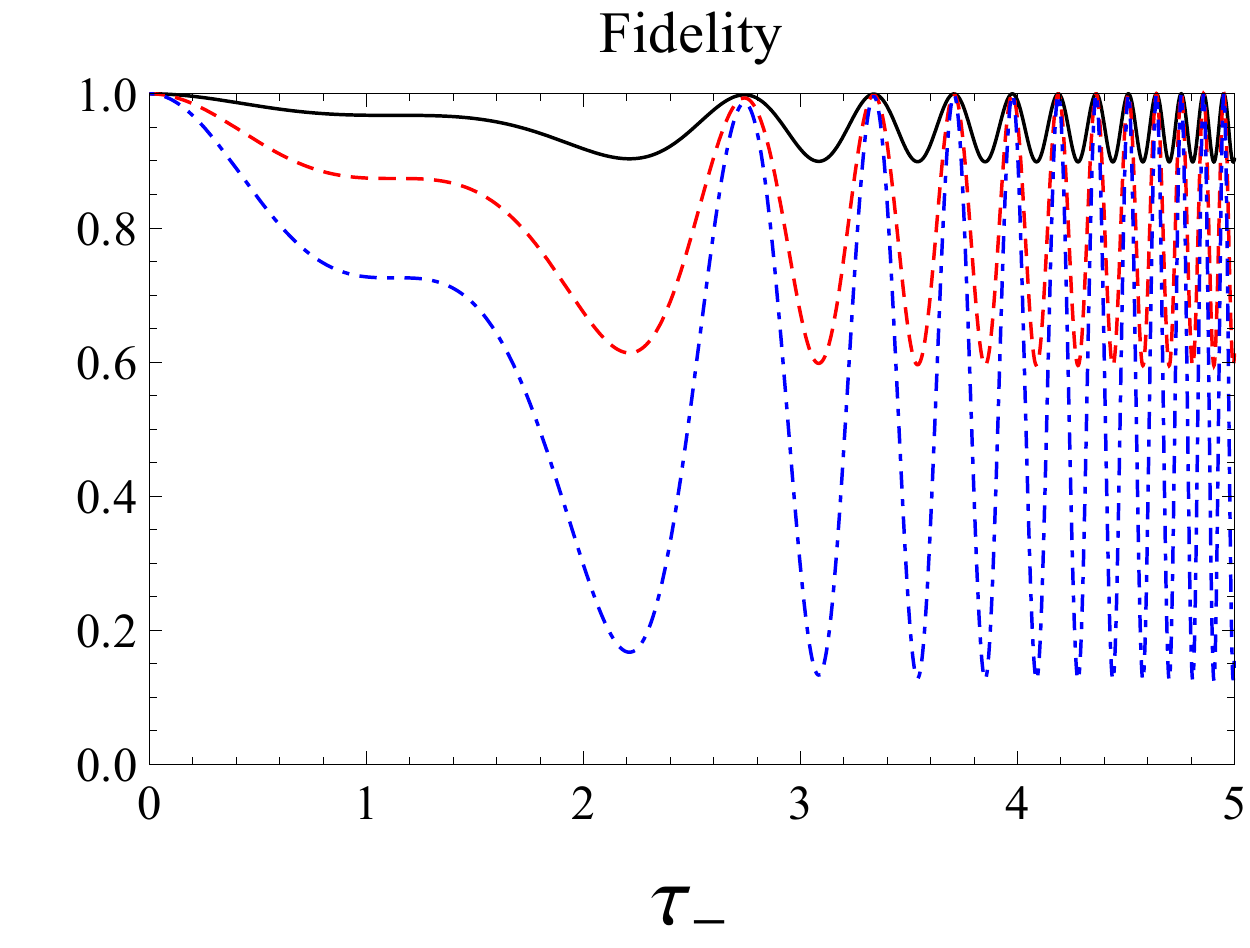}
\caption{Fidelity versus dimensionless time $\tau_-={|\Gamma_-| \over \hbar}\, t $  when the initial state is the Werner state in Equation (\ref{st-werner}) for $\alpha =0.25$ (black, solid), $\alpha =0.55$ (red, dashed), and $\alpha =0.9 $ (blue, dot-dashed) in Case~1---\textbf{left} and Case 2---\textbf{right}.}
\label{fig-mixta-werner-case1-2}
\end{figure}

Finally, in Figure~\ref{fig-mixta-werner-case1-2} the fidelity between the Werner state and the evolved Werner state is reported versus the dimensionless time $\tau_-$ for different values of $\alpha$.
The analytical derivation of the expression of the fidelity between the Werner state and the generalized Werner state is reported in Appendix~\ref{Fid W States} by Equation (\ref{fidelit-familie}). Furthermore, one replaces the parameters $\mu $ and $\nu $ of the generalized Werner state by $c_{01}(t)$ and $c_{10}(t)$, respectively, according to their expressions (\ref{c01}) and (\ref{c10}) in order to obtain the analytical expression of the fidelity between the Werner state and the desired state, i.e., the evolved Werner state.

We see that the curves of Figure~\ref{fig-mixta-werner-case1-2} exhibit a time behavior qualitatively similar to the ones related to the pure state $\ket{\Psi^-}$.
The physical reason lies on the fact that as pointed out before, the time evolution of the Werner state, according to the Hamiltonian model under scrutiny, is traceable back by the time evolution of the state $\ket{\Psi^-}$.
The parameter $\alpha$ practically scales the curves as it happens for the fidelity in Figure~\ref{fig-mixta-werner-case1-2} as well as for the concurrence in Figure~\ref{fig-werner-case2-plan}.

\section{Conclusions}
\label{sec-concl}
In this paper, we have investigated the emergence and the time behavior of the quantum correlations generated in a driven system of two interacting spin-1/2's subjected to local time-dependent magnetic fields.
To this end, we have studied the evolution of the concurrence and the quantum discord when the system is acted upon by specific fields for which the exact time evolution operator is known~\cite{Messina-2016}. The~specific time-dependent (controlled) scenarios we analyzed are based on the capability of generating a sech pulse. The~problem of a single spin subjected to a sech pulse dates back to the 1930s and has been formulated and treated by Rosen and Zener~\cite{Rosen}. Since the experimental setup for such a pulse turns out to be easily realizable~\cite{Economu,Greilich,Poem}, even today it is still of theoretical and applicative interest and appears indeed in many-spin Hamiltonian models~\cite{Hioe,Kyoseva,Vitanov}. The~application of inhomogeneous and time-dependent magnetic fields on a pair of coupled spins exploits the
so-called Scanning Tunneling Microscopy (STM)~\cite{Khajetoorians,Yan,Bryant,Tao,Lutz,Wieser,Sivkov}. The~exchange interaction between the spin on the tip of the Microscope and the spin of interest in the pair origins the local and desired magnetic field. The~geometrical relative configuration between the tip and the target spin is adjustable enabling, at least in principle,
the generation of effective local time-dependent magnetic fields at will.
It is of relevance moreover to emphasize that even if the exact treatment of the quantum dynamics of time-dependent Hamiltonian models are rare, our ability to find the evolution operator is not limited to the cases we have reported~\cite{MN}. The~two scenarios selected in this paper are exemplary ones since they are non-trivial, exactly treatable as well as within the experimental reach.

The symmetry properties of our time-dependent Hamiltonian model play a crucial role since it guarantees that an initial $X$ density matrix evolves keeping such a structure at any time instant and, on the other hand, that the quantum discord of such a state could be analytically determined~\cite{Roberto}.
This is why we choose an $X$ state as initial condition and in this class we concentrate on generic $\alpha$-parametric Werner states.

Our analysis exactly predicts in both time-dependent scenarios the presence of sudden death-sudden revival phenomena in the concurrence as well as a non-vanishing quantum discord.
Many papers deal with the same issue, but mainly focusing on open quantum systems where death and rebirth of entanglement stems instead from the interaction with the surrounding
environment~\cite{Mazzola,Namitha,Bahari,Shaukat}.
We emphasize that our prediction of the zero-concurrence plateaux is based on the knowledge of the structure of the class of the extended Werner states $\eta^{(\alpha)}_{\mu,\nu}$, which enables a transparent distinction between domains of zero concurrence and domains of non-vanishing concurrence in the $\alpha$-$\mu$ parameter space as illustrated in Figures~\ref{fig-alfa-conc-zero} and~\ref{fig-conc-density-plot}.

Comparing the two plots in Figure~\ref{fig-conc-werner-case-pura}, we finally notice a peculiar difference in the asymptotic behavior of concurrence and quantum discord in the two controlled scenarios investigated in this paper.
We succeeded in interpreting the dynamical origin of such a difference evaluating the time behavior of the fidelity of the initial Werner state with respect to the evolved one.

A possible perspective of the present work could consist of studying the same two-spin system in the presence of a quantum harmonic oscillator bath making in this way more realistic the physical scenario.
The quantum dynamics of this open quantum system could be treated with the Feshbach approach leading to the consideration of appropriate effective non-Hermitian Hamiltonians~\cite{Feshbach1,Feshbach2} or, alternatively, it could be based on the partial Wigner transpose approach~\cite{SHGM}.

\vspace{1cm}
{\bf Author Contributions:} I.G. wrote the main part of the manuscript and is responsible for the original draft preparation and the plots present in the paper; R.G., T.M., A.I. and A.M. are responsible for writing, reviewing and editing; A.M. and A.I. are responsible for the supervision of the project. All authors have read and agreed to the published version of the manuscript.

\section*{Acknowledgments}
The work of I.G. was supported by the funding agency CNCS-UEFISCDI of the Romanian Ministry of Research and Innovation through grant PN-III-P4-ID-PCE-2016-0794. A.I. acknowledges the financial support received from the Romanian Ministry of Education and Research, through the Project PN 19 06 01 01/2019. R.G. acknowledges support by research funds difc 3100050001d08+, University of Palermo, in memory of Francesca~Palumbo.

\appendix
\begin{appendix}

\section{Analytical Solutions of the Hamiltonian Model Given in Section~\ref{hamilt-model}}
\label{exact-solutions}

In this Appendix we provide some results that were obtained in Ref.~\cite{Messina-2016}. Let us define:
\begin{equation}
 \phi_{\Gamma_\pm}:=-\arctan \left[ {\pm \gamma_{12} + \gamma_{21} \over \gamma_{11} \mp \gamma_{22} } \right].
 \label{fi-gama}
\end{equation}

Case 1. If the two magnetic fields vary over time as follows
\[
\hbar \omega_{A,B}(t) = \frac{|\Gamma_+|}{\cosh(2\tau_+)} \pm \frac{|\Gamma_-|}{\cosh(2\tau_-)},
\]
then the solutions for the two fictitious spin-1/2 particles are (see Equation (\ref{unit-U})):
\begin{eqnarray}
|a_+(t)| &=& \sqrt{\frac{ \cosh(2\tau_+) + 1 }{2 \cosh(2\tau_+)}}, \qquad \qquad \; \; \; \;
 |b_+(t)| = \sqrt{\frac{ \cosh(2\tau_+) - 1 }{2 \cosh(2\tau_+)}}, \label{a4}\\
 \phi_{a}^+(t) & =& - \arctan [ \tanh ( \tau_+ ) ] - \tau_+, \qquad
 \phi_b^+(t) = \phi_{\Gamma_+} - \arctan [ \tanh ( \tau_+ ) ] + \tau_+ - {\pi \over 2}, \label{a5}\\
|a_-(t)|& =& \sqrt{\frac{ \cosh(2\tau_-) + 1 }{2 \cosh(2\tau_-)}}, \qquad \qquad \; \; \; \;
 |b_-(t)| = \sqrt{\frac{ \cosh(2\tau_-) - 1 }{2 \cosh(2\tau_-)}}, \label{a6}\\
\phi_{a}^-(t) &=& - \arctan [ \tanh ( \tau_- ) ] - \tau_-, \qquad
 \phi_b^-(t) = \phi_{\Gamma_-} - \arctan [ \tanh ( \tau_- ) ] + \tau_- - {\pi \over 2}.\label{a7}
 \label{a and b case 1}
\end{eqnarray}

Case 2. Likewise, if the two local magnetic fields change in time as
\[
\hbar \omega_{A,B}(t) = \frac{|\Gamma_+|}{\cosh(2\tau_+)} \pm
{|\Gamma_-| \over 4} \biggl[ { 3 \over \cosh(\tau_-) } - \cosh(\tau_-) \biggr], \\
\]
then the solutions, in this case, read
\begingroup\makeatletter\def\f@size{9}\check@mathfonts
\def\maketag@@@#1{\hbox{\m@th\normalsize\normalfont#1}}%
\begin{eqnarray}
|a_+(t)|~ =~ \sqrt{\frac{ \cosh(2\tau_+) + 1 }{2 \cosh(2\tau_+)}}, \qquad \qquad \qquad \qquad \; \;
 |b_+(t)| = \sqrt{\frac{ \cosh(2\tau_+) - 1 }{2 \cosh(2\tau_+)}}, \label{a9}\\
 \phi_{a}^+(t) = - \arctan [ \tanh ( \tau_+ ) ] - \tau_+, \qquad \qquad \; \; \; \; \; \;
 \phi_b^+(t) = \phi_{\Gamma_+} - \arctan [ \tanh ( \tau_+ ) ] + \tau_+ - {\pi \over 2}, \label{a10}\\
 |a_-(t)|~ =~ \frac{1}{\cosh(\tau_-)}, \qquad \qquad \qquad \qquad \qquad \; \; \; \; \; \;
 |b_-(t)| = \tanh(\tau_-), \label{a11}\\
 \phi_{a}^-(t)~=~ - \arctan \Bigl[ \tanh \Bigl( {\tau_- \over 2} \Bigr) \Bigr] - {1 \over 2} \sinh (\tau_-), \; \; \;
  \phi_b^-(t) = \phi_{\Gamma_-} - \arctan \Bigl[ \tanh \Bigl( {\tau_- \over 2} \Bigr) \Bigr] + {1 \over 2} \sinh (\tau_-) - {\pi \over 2}. \label{a12}
\end{eqnarray}
\endgroup

\section{Quantum Discord for $X$ States}
\label{sec-discord}

Quantum discord is an important tool for measuring quantum correlations in a bipartite quantum system, which was defined by Olivier and Zurek~\cite{Zurek}, being based on the non-equivalence in the quantum case of two classical definitions of the mutual information.  The classical mutual information can be defined in two ways:
\begin{eqnarray}
I(A:B)&=&H(A)+H(B)-H(A,B); \label{def1}\\
J(A:B)&=&H(A)-H(A|B), \label{def2}
\end{eqnarray}
where $H$ is the Shannon entropy, while $H(A|B)$ is the conditional Shannon entropy. In the classical case, the definitions (\ref{def1}) and (\ref{def2}) coincide.

For the generalization to the quantum case, one must consider a bipartite state $\rho_{AB}$~\cite{Zurek}:
\begin{equation}
{\cal I}(\rho_{AB})=S(\rho_A)+S(\rho_B)-S(\rho_{AB}),
\label{q-mut-inf}
\end{equation}
with $S(\rho )$ being the von Neumann entropy $S(\rho)=-\mbox{Tr}(\rho\log_2\rho)$, while $\rho_{A(B)}:=\mbox{Tr}_{B(A)}\, \left(\rho_{AB}\right)$ denote the reduced states of the two subsystems. Equation (\ref{q-mut-inf}) represents the quantum mutual information between the two subsystems, A and B.
The quantum analogue of the formula (\ref{def2}) is more complicated and depends on the von Neumann measurements made on the second system $B$. Let us denote by $\{ \Pi^B_k \}$ the set of one-dimensional projectors performed on the system $B$. The~final state of the system $A$, after the measurement on the system $B$ led to the outcome $j$ is~\cite{Zurek}:
\begin{equation}
\rho_{A|\Pi^B_j}=\frac{1}{p_j}\, \mbox{Tr}_B(I\otimes \Pi^B_j\, \rho_{AB}\, I\otimes \Pi^B_j),
\end{equation}
where the probability is given by $p_j=\mbox{Tr}(\rho_{AB}\, I\otimes \Pi^B_j)$. The~quantum conditional entropy is obtained by considering all the possible outcomes:
\begin{equation}
S(\rho_{A|\{ \Pi^B_j\}})=\sum_jp_j\, S(\rho_{A|\Pi^B_j}).
\end{equation}

The associated quantum mutual information, which generalizes Equation (\ref{def2}), is
\begin{equation}
{\cal J}(\rho_{AB})|_{\{\Pi^B_j\}}=S(\rho_A)-S(\rho_{A|\{ \Pi^B_j\}}).
\end{equation}

In Refs.~\cite{Zurek,Henderson} the classical correlation is defined by considering the supremum over all the possible von Neumann measurements $\{ \Pi^B_k \}$:
\begin{equation}
{\cal C}(\rho_{AB})=\sup_{\{\Pi^B_j\}}{\cal J}(\rho_{AB})|_{\{\Pi^B_j\}}.
\end{equation}

The quantum $A$-discord is defined as~\cite{Zurek}:
\begin{equation}
D_A(\rho_{AB})={\cal I}(\rho_{AB})-{\cal C}(\rho_{AB}),
\label{discord}
\end{equation}
being a measure of quantum correlations of a two-party quantum state. In addition, there is a second definition, namely the quantum $B$-discord, which considers the von Neumann measurements performed on the first system. The~quantum discord is asymmetric under the change $A\leftrightarrow B$, i.e., it depends on which subsystem the measurements are made on. In this paper, we use the quantum $A$-discord, which we will call it briefly $D(\rho)=D_A(\rho)$.

We have shown in Section~\ref{hamilt-model} that if one starts with a two-qubit system found in an $X$ state, then the evolved state under the influence of the two external magnetic fields is also an $X$ state described by $\rho_{\mbox x}(t)$. If the input state is the Werner state, then the elements of the density operators $\rho_{\mbox x}(t)$ are given by Equations (\ref{st-x-timp}), whose Fano parametrization is given by Equation (\ref{fano-x}), i.e., being described by a non-diagonal matrix $T$.
By applying the local unitary transformation (\ref{op-unit-loc}), one can bring the state $\rho_{\mbox x}(t)$ to its canonical form given by Equation (\ref{x-can-fano}), as we have discussed in Section~\ref{canonic}. Furthermore, we use the algorithm found by Li~et~al. for evaluating the quantum discord of an $X$ state in the canonical form~\cite{Li}. First, we compute the parameters $r(t)$, $s(t)$, $c_1(t)$, $c_2(t)$, $c_3(t)$ by using Equations (\ref{fano-can}) and~(\ref{st-x-timp}).
The eigenvalues of the canonical $X$ state (\ref{x-can-fano}) are given by~\cite{Li}:
\begin{eqnarray}
\lambda_{1,2} &=&\frac{1}{4}\left[ 1-c_3\pm\sqrt{(r-s)^2+(c_1+c_2)^2} \right],\nonumber \\
\lambda_{3,4} &=&\frac{1}{4}\left[ 1+c_3\pm\sqrt{(r+s)^2+(c_1-c_2)^2} \right]. \label{val-pr}
\end{eqnarray}

Let us define the monotonically decreasing function $u$ for $x\in[0, 1]$:
\begin{equation}
u(x)=-\frac{1-x}{2}\, \log_2(1-x)-\frac{1+x}{2}\, \log_2(1+x).
\end{equation}

One can easily evaluate the von Neumann entropy of the reduced density operators:
$
S(\rho_A)=1+u(r),\; \;
S(\rho_B)=1+u(s).
$ According to Equation (\ref{q-mut-inf}), the quantum mutual information has the~expression:
\begin{equation}
{\cal I}(\rho_{AB})=2+ u(r) + u(s) + \sum_{j=1}^4\, \lambda_j\, \log_2\, \lambda_j.
\end{equation}

Let us consider the following functions~\cite{Li}:
\begin{eqnarray}
f_1 &=& -\frac{1+r+s+c_3}{4}\log_2\frac{1+r+s+c_3}{2(1+s)}
   -\frac{1-r+s-c_3}{4}\log_2\frac{1-r+s-c_3}{2(1+s)}  \nonumber\\
     &&-\frac{1+r-s-c_3}{4}\log_2\frac{1+r-s-c_3}{2(1-s)}
      -\frac{1-r-s+c_3}{4}\log_2\frac{1-r-s+c_3}{2(1-s)}, \nonumber \\
f_2 &=& 1+u\left( \sqrt{r^2+c_1^2}\right), \; \; \;
f_3 = 1+u\left( \sqrt{r^2+c_2^2}\right). \nonumber
\end{eqnarray}

Li~et~al. have proved that the classical correlation has the expression~\cite{Li}:
\begin{equation}
{\cal C}(\rho_{AB})= S(\rho_A) - \min\{f_1, f_2, f_3\}.
\end{equation}

Therefore, the quantum discord can be evaluated by using Eq. (\ref{discord}).

\section{Comparison between Concurrence and Quantum Discord of the Werner State Subjected to Constant Magnetic Fields }
\label{constant}

Let us now investigate the case of constant applied fields, which means that both $\omega_A$ and $\omega_B$ are time-independent. According to Equation (\ref{omega-gama}), one obtains that also $\Omega_\pm $ are constant. In addition, we consider $\gamma_{11}=\gamma_{22}=c$ and $\gamma_{12}=\gamma_{21}$.  By using Equation (\ref{omega-gama}), one obtains the expression of $\Gamma_-$, i.e., $\Gamma_-=2 \, c$. Therefore, we get:
\begin{eqnarray}
a_\pm(t) &=& e^{\mp i\gamma_{33}t/\hbar} \left[ \cos(\tau_\pm) - i {\Omega_\pm \over \nu_\pm} \sin(\tau_\pm) \right],\nonumber \\
b_\pm &=& -ie^{\mp i\gamma_{33}t/\hbar} {\Gamma_\pm \over \nu_\pm} \sin(\tau_\pm), \nonumber
\end{eqnarray}
where we have denoted $\nu_\pm \equiv \sqrt{\Omega_\pm^2+|\Gamma_\pm|^2}$,
$\tau_-=\nu_- t/\hbar $. With the notation $\Omega_-=\beta \, c$, one obtains:
\[
\frac{\Omega_-}{\nu_-}=\frac{\beta }{\sqrt{\beta^2+4}}, \; \; \;
\frac{\Gamma_-}{\nu_-}=\frac{2 }{\sqrt{\beta^2+4}}.
\]

According to Equation (\ref{conc-X}), the expression of the concurrence in the case of constant fields~becomes:
\begin{equation}
C=\max \left\{  0, \alpha \, \sqrt{1-\frac{16\beta^2}{(\beta^2+4)^2\, \sin^4 (\tau_-)}}-\frac{1-\alpha }{2} \right\}.
\end{equation}

\begin{figure}
\centering
\includegraphics[width=5cm]{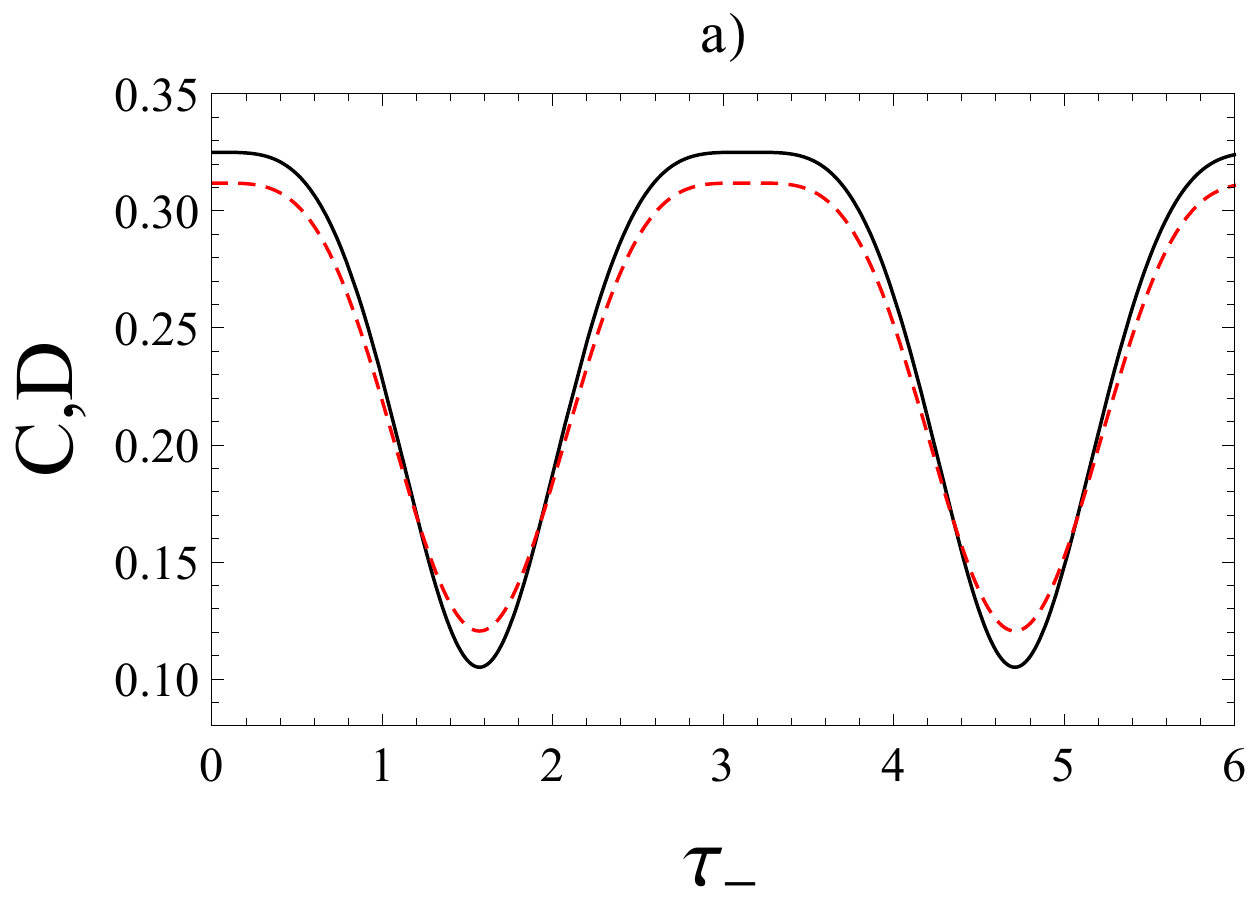}
\includegraphics[width=5cm]{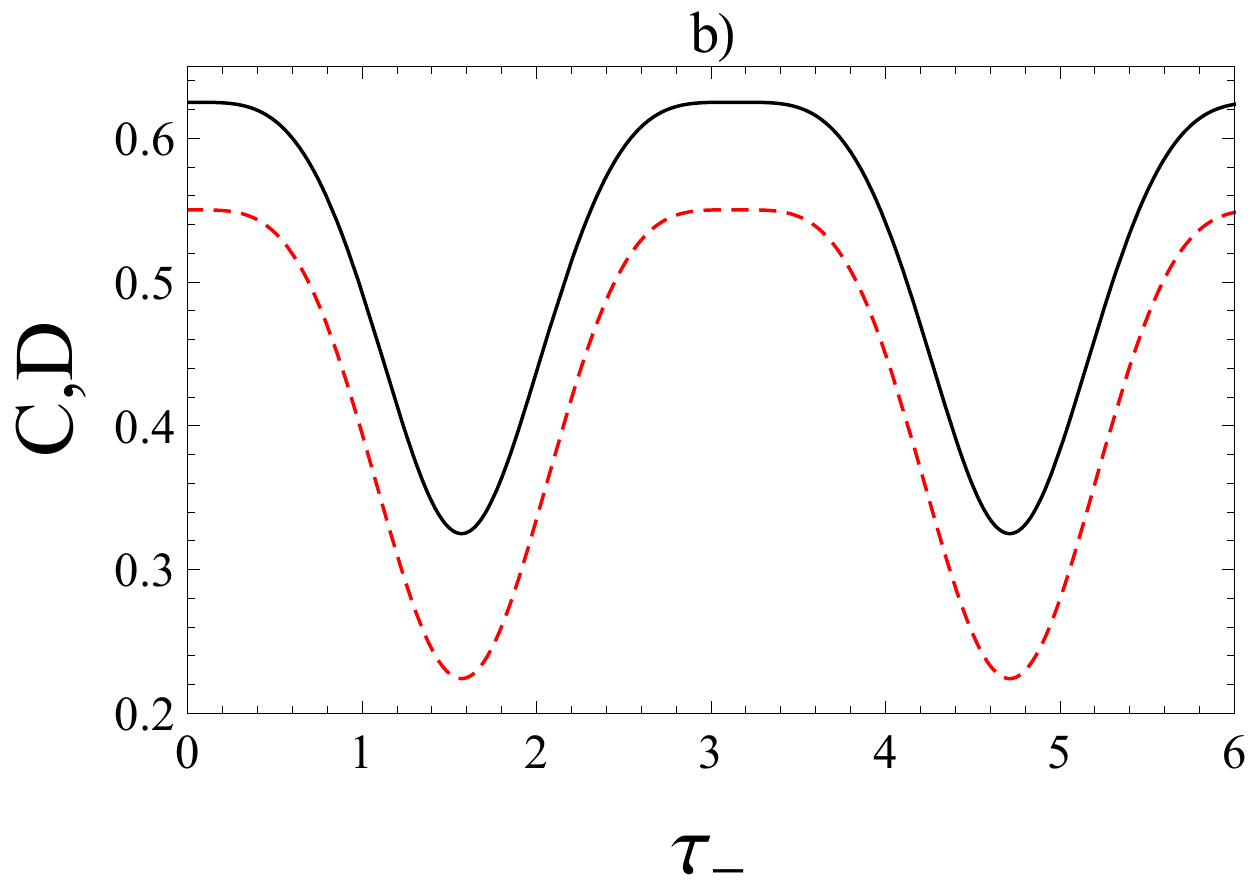}
\includegraphics[width=5cm]{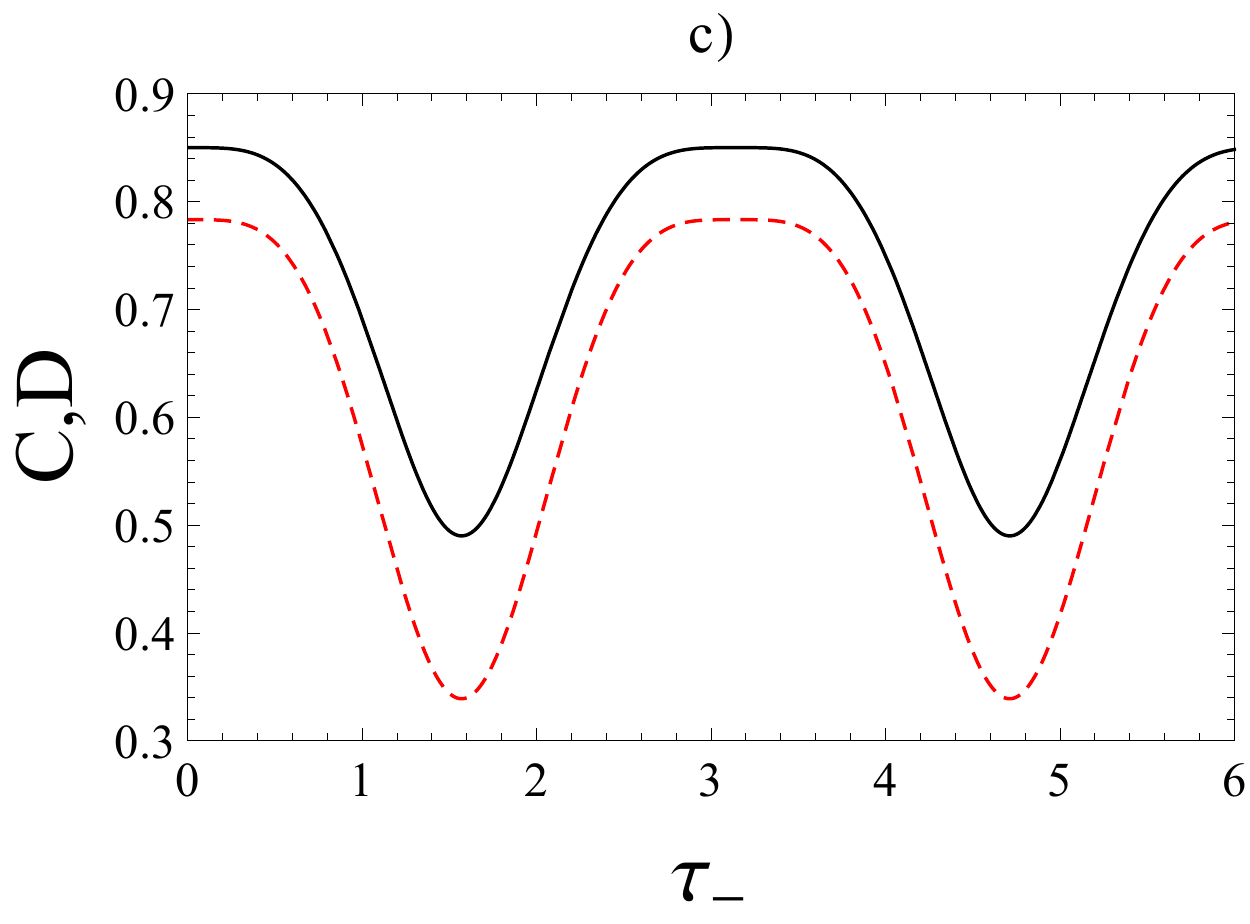}
\caption{Concurrence (black, solid) and quantum discord (red, dashed) when the initial state is the Werner state (\ref{st-werner}) in the case of constant fields, when $\beta = 1$, for: (\textbf{a}) $\alpha =0.55 $, (\textbf{b}) $\alpha =0.75$, (\textbf{c}) $\alpha = 0.9$.}
\label{const-1}
\end{figure}

\begin{figure}
\centering
\includegraphics[width=5cm]{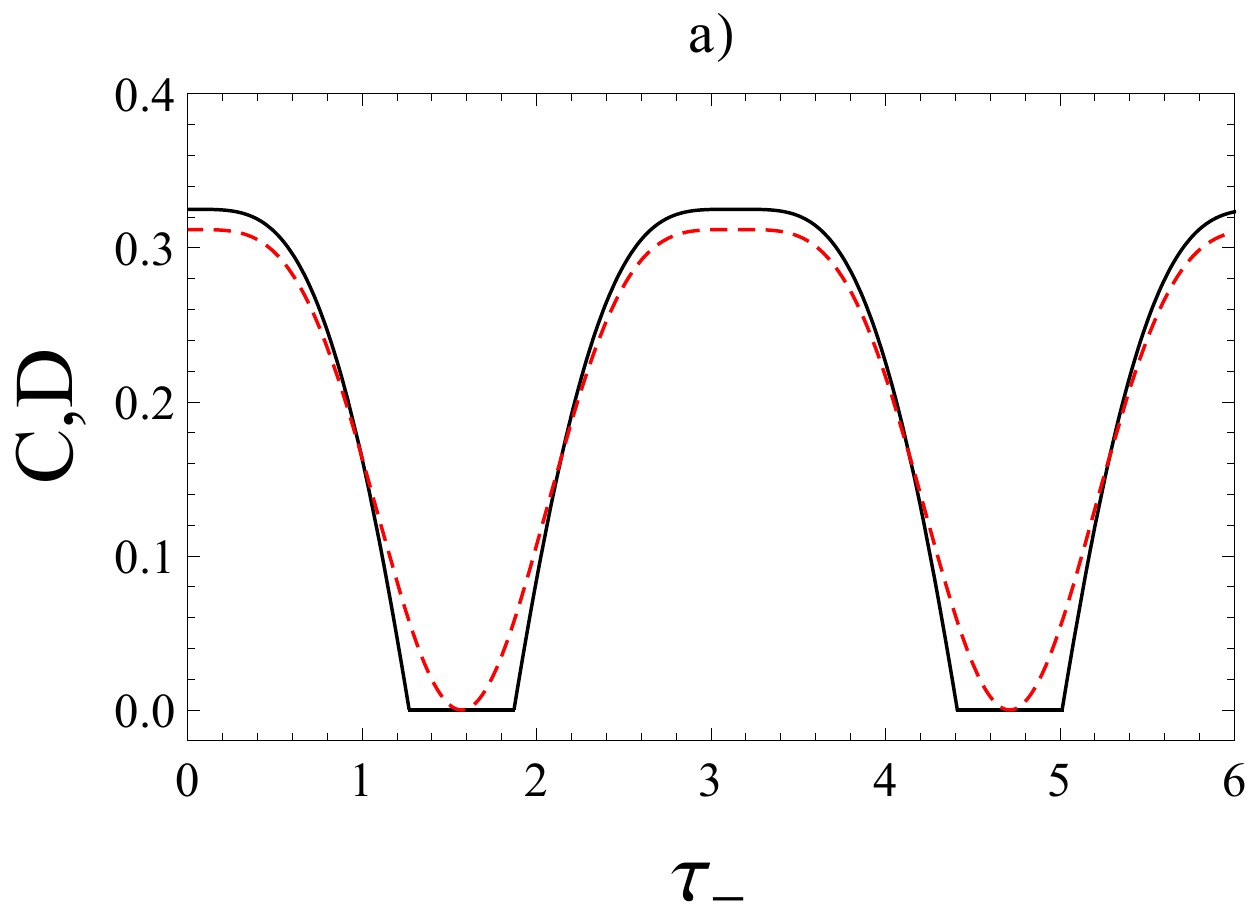}
\includegraphics[width=5cm]{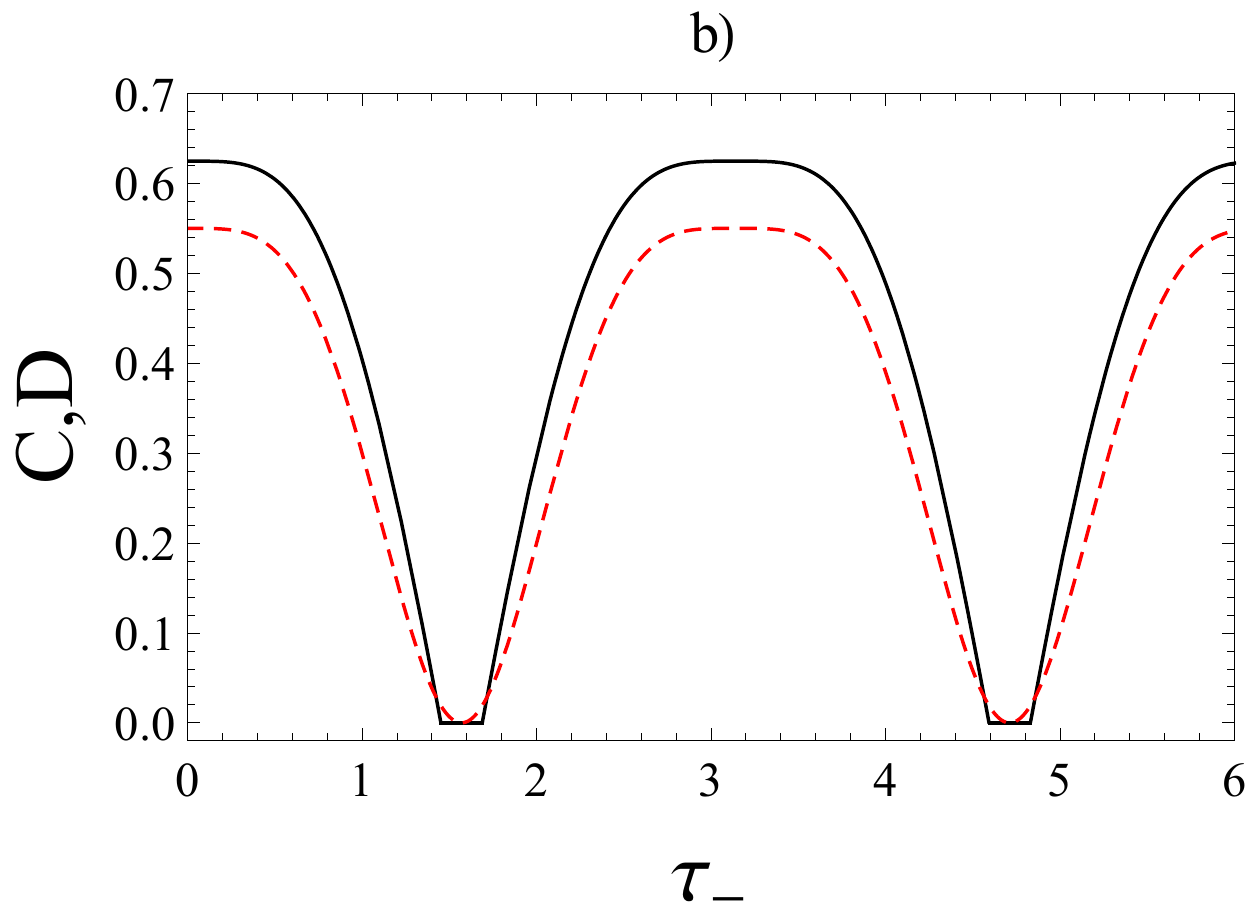}
\includegraphics[width=5cm]{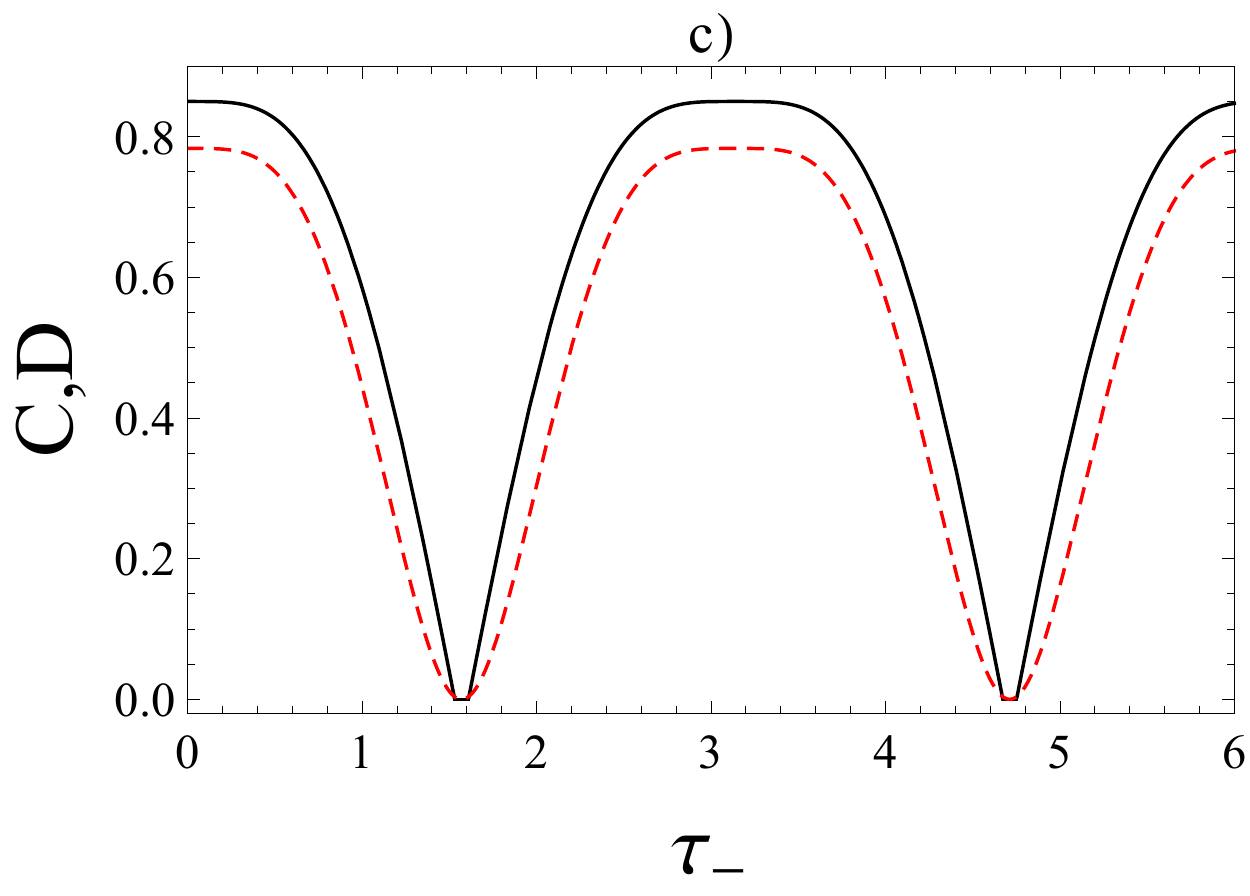}
\caption{Concurrence (black, solid) and quantum discord (red, dashed) when the initial state is the Werner state (\ref{st-werner}) in the case of constant fields, when $\beta = 2$, for: (\textbf{a}) $\alpha =0.55 $, (\textbf{b}) $\alpha =0.75$, (\textbf{c}) $\alpha = 0.9$.}
\label{const-2}
\end{figure}

Concurrence and quantum discord for the time-independent case are reported in Figures~\ref{const-1} and~\ref{const-2}. We remark that both the parameter $\alpha$ and the parameter $\beta$ influence the time behavior of both the concurrence and the discord. In particular, we see that the parameter $\alpha$ sets the range of variation of the curves; precisely, when $\alpha \to 1$ both the two quantities are closer to higher values, as it can be appreciated also from Figures~\ref{fig-werner-case1-plan} and~\ref{fig-werner-case2-plan}. The~parameter $\beta$, instead, qualitatively determines the shape of these curves and it is responsible for the presence of possible plateaux of the concurrence, as Figure~\ref{const-2} clearly~shows.

\unskip

\section{Fidelity between the Werner State and the Generalized Werner State}
\label{Fid W States}

In the following we are interested in computing the fidelity between the Werner state and the evolved state $\rho(t)$. We have proved in Section~\ref{dynamics} A that the state $\rho(t)$  is a particular state of $\eta_{\mu ,\nu }^{(\alpha )}$ of Equation (\ref{st-gen-eta}),  obtained for the particular case $\mu =c_{01}(t)$ and $\nu =c_{10}(t)$.

The fidelity between two mixed states $\rho_1$ and $\rho_2$ is defined by~\cite{Jozsa}:
\begin{equation}
{\cal F}(\rho_1,\rho_2):=\left[\mbox{Tr}
\sqrt{\rho_1^{1/2}\rho_2\,\rho_1^{1/2}}\right]^2.
\end{equation}

Further we evaluate the fidelity between the Werner state $\rho_W$ and the state $\eta_{\mu ,\nu }^{(\alpha )}$. Let us denote by $P$, $Q$, $R$ the following parameters:
\begin{eqnarray}
P&=&\frac{1}{8}\, (1-\alpha^2)+\frac{1}{8}\, \alpha \, (1-\alpha )\, |\mu +\nu |^2 + \frac{1}{8}\, \alpha \, (1+3\, \alpha )\, |\mu -\nu |^2 ,\nonumber \\
Q&=&\frac{1}{256}\, (1-\alpha)^3(1+3\, \alpha)+\frac{1}{128}\, \alpha \, (1+3\alpha)(1-\alpha)^2 \left(|\mu +\nu |^2+|\mu -\nu |^2 \right). \nonumber
\end{eqnarray}

The eigenvalues of $\rho_W^{1/2}\eta_{\mu ,\nu }^{(\alpha )} \,\rho_W^{1/2}$ are:
\begin{eqnarray}
\zeta_1&=&\zeta_2=\frac{(1-\alpha)^2}{16}; \nonumber \\
\zeta_3&=&\frac{1}{2}\left( P+\sqrt{P^2-4\, Q}\right); \nonumber \\
\zeta_4&=&\frac{1}{2}\left( P-\sqrt{P^2-4\, Q}\right).\nonumber
\end{eqnarray}

Therefore, the fidelity has the expression
\begin{equation}
{\cal F}(\rho_W,\eta_{\mu ,\nu }^{(\alpha )})=\left( \frac{1-\alpha}{2}+\sqrt{\zeta_3}+ \sqrt{\zeta_4}\right)^2.
\label{fidelit-familie}
\end{equation}

\end{appendix}

\end{document}